\documentclass[twocolumn]{aastex63}

\usepackage{graphicx}
\usepackage{amsmath}	
\usepackage{amssymb}	
\usepackage{mathrsfs}

\usepackage{bigints}

\newcommand{\msun}{M_\odot}

\newcommand{\jlw}{J_{\rm LW}}
\newcommand{\vbsm}{v_{\rm bsm}}
\newcommand{\HH}{\mathrm{H_2}}
\newcommand{\fesc}{f_{\mathrm{esc}, \mathrm{LW}}}
\newcommand{\msunyr}{M_\odot~{\rm yr}^{-1}}
\newcommand{\gcc}{\rm g~{cm}^{-3}}
\newcommand{\Mdot}{\dot{M}_{\star}}
\newcommand{\K}{{\rm K}}
\newcommand{\cc}{{\rm cm}^{-3}}

\usepackage{CJK}
\usepackage{amsmath}
\usepackage{xcolor} 
\definecolor{bg-green}{rgb}{0.8588,0.9333,0.8666}
\definecolor{Q-color}{rgb}{0.5,0.1,0.5}

\shorttitle{}
\shortauthors{Li et al.}

\begin{document}

\begin{CJK*}{UTF8}{gbsn}

\title{Evolution of high-redshift quasar hosts and promotion of massive black hole seed formation
}

\correspondingauthor{Kohei Inayoshi}
\email{inayoshi@pku.edu.cn}

\author[0000-0002-1044-4081]{Wenxiu Li (李文秀)}
\affiliation{Kavli Institute for Astronomy and Astrophysics, Peking University, Beijing 100871, China}

\author[0000-0001-9840-4959]{Kohei Inayoshi}
\affiliation{Kavli Institute for Astronomy and Astrophysics, Peking University, Beijing 100871, China}

\author[0000-0002-6164-8463]{Yu Qiu (邱宇)}
\affiliation{Kavli Institute for Astronomy and Astrophysics, Peking University, Beijing 100871, China}

\begin{abstract}

High-redshift luminous quasars powered by accreting supermassive black holes (SMBHs) with mass $\gtrsim 10^9~\msun$ constrain their formation pathways. 
We investigate the formation of heavy seeds of SMBHs through gas collapse in the quasar host progenitors,
using merger trees to trace the halo growth in highly-biased, overdense regions of the universe.
The progenitor halos are likely irradiated by intense H$_2$-photodissociating radiation from nearby star-forming galaxies and heat the interior gas by successive mergers.
The kinetic energy of the gas originating from mergers as well as baryonic streaming motion prevents gas collapse and delays prior star formation.
With a streaming velocity higher than the root-mean-square value, gas clouds in nearly all $10^4$ realizations of merger trees enter the atomic-cooling stage and begin to collapse isothermally with $T \simeq 8000 ~\K$ via Ly$\alpha$ cooling.
The fraction of trees which host isothermal gas collapse is $14\%$ and increases with streaming velocity, while the rest form H$_2$-cooled cores after short isothermal phases.
If the collapsing gas is enriched to $Z_{\rm crit}\sim 2\times 10^{-3}~Z_\odot$,
requiring efficient metal mixing, this fraction could be reduced by additional cooling via metal fine-structure lines.
In the massive collapsing gas, the accretion rate onto a newly-born protostar ranges between $3\times10^{-3}-5~\msunyr$,
among which a large fraction exceeds the critical rate suppressing stellar radiative feedback.
As a result, we expect a distribution of stellar mass (presumably BH mass) ranging from several hundred to above $10^5~\msun$,
potentially forming massive BH binary mergers and yielding gravitational wave events.
\end{abstract}

\keywords{Supermassive black holes (1663); Quasars (1319); High-redshift galaxies (734)}


\vspace{5mm}
\section{Introduction}
\label{sec:intro}

Supermassive black holes (SMBHs) with masses of $10^{6-9}~\msun$ are one of the most fundamental ingredients
on the structure formation paradigm and are believed to coevolve with their host galaxies over the cosmic timescale
through gas feeding and feedback processes \citep{2013ARA&A..51..511K}.
The existence of high-redshift quasars at $z\ga 6$ suggests that such monster SMBHs form in the first billion years of the cosmic age
\citep{2006AJ....131.1203F,2011Natur.474..616M,2015Natur.518..512W,Jiang_2016,2018ApJS..237....5M,Onoue_2019,2021ApJ...907L...1W}
via rapid assembly processes, such as the formation of heavy BH seeds (initial mass), rapid mass growth via gas accretion, or 
a combination of the two mechanisms (see a review by \citealt{2020ARA&A..58...27I}).

For massive seed BH formation, a sufficiently high accretion rate of gas onto stellar objects is required.
In early protogalaxies where the halo virial temperature is as high as $T_{\rm vir}\simeq 10^4~\K$ and
the temperature of a self-gravitating gas cloud is as warm as that value,
the mass accretion rate is expected to be $\dot{M}\simeq c_{\rm s}^3/G \simeq 0.1~\msunyr (T/10^4~\K)^{3/2}$,
where $c_{\rm s}$ is the sound speed of the gas and $G$ is the gravitational constant.
To keep the gas warm against efficient cooling via H$_2$ lines, several mechanisms suppressing, delaying, and counteracting H$_2$ formation/cooling
have been proposed by many previous studies in literature:
photo-dissociation of H$_2$ by Lyman-Werner (LW) radiation \citep{Omukai2001,Oh_Haiman2002,Shang2010,Latif2013,Inayoshi_Omukai_Tasker2014,Sugimura2014,
Regan2014,2014MNRAS.445.1056V,Chon_2016},
supersonic baryonic streaming motion relative to dark matter \citep{2014MNRAS.439.1092T,Hirano18,Inayoshi_Li_Haiman2018,2019MNRAS.484.3510S}, and 
rapid halo mergers which cause heating \citep{Yoshida2003,2019Natur.566...85W,Lupi2021} as well as reduce H$_2$ cooling through accretion shocks \citep{2014MNRAS.439.3798F}
All the three processes bring the gas cloud into a dense and hot region on the gas phase diagram,
where collisional dissociation from the excited rovibrational levels of H$_2$ reduces the H$_2$ fraction \citep{2012MNRAS.422.2539I}.
In the subsequent stage, the gas collapses almost isothermally, keeping itself as warm as $T\simeq 3000-8000~\K$ and 
avoiding vigorous gas fragmentation into smaller clumps \citep{2003ApJ...596...34B,Latif2013, Inayoshi_Omukai_Tasker2014,2015MNRAS.446.2380B,Chon_2018}.
Due to global and monolithic collapse of the warm cloud, the embryonic protostar is fed by rapidly accreting gas
at a rate of $\ga 0.1~\msunyr$ through a compact accretion disk where gas clumps could quickly migrate inward and merge with the central protostar \citep{2014MNRAS.445.1549I,2016MNRAS.459.1137S}.
Moreover, since the protostar evolves with an expanding stellar envelope due to rapid entropy inject from the accreting matter,
the surface temperature is limited to $T_{\rm eff}\simeq 5000~\K$, which is too low for the protostar to emit ionizing radiation \citep{Hosokawa2013,Haemmerle2018}.
As a result of inefficient radiative feedback, the protostar would likely reach $\sim 10^{5-6}~\msun$ before the end of its lifetime
and collapse into a massive seed BH.
However, those formation sites of mass seed BHs are expected to be as rare as the number density of high-$z$ quasars in a comoving volume
\citep[$n_{\rm SMBH}\sim 1-10~{\rm Gpc}^{-3}$ from ][]{2010AJ....139..906W}.

Recent cosmological hydrodynamical simulations have suggested that the conditions required to form massive seeds
should be more modest than previously considered \citep[e.g.,][]{2019Natur.566...85W}.
Even with a moderate level of LW radiation, streaming motion and merger heating,
a high mass accretion rate is sustained at larger radii in a protogalaxy,
although the isothermality of gas is not maintained at high densities ($n\ga 100~\cc$).
Under such less stringent situations, the average mass accretion rate onto the central protostar is reduced but the peak rate 
can exceed the critical rate for bifurcating the protostellar evolution \citep{2015MNRAS.452.1026L,2017Sci...357.1375H,2020OJAp....3E..15R}.
As a result, the central star grows to the intermediate mass regime at $M_\star \simeq 100-10^4~\msun$, 
which is lower than originally expected the expected mass for a SMS but still massive enough to form 
massive seeds that will end up as high-$z$ SMBHs \citep[][Toyouchi et al. in prep]{2020MNRAS.499.5960S}.
Therefore, those environmental effects are potentially important to initiate intermediate massive BHs (IMBHs) in the high-$z$ universe by $z\sim~6-7$ \citep{2020ARA&A..58...27I},
and form gravitational-wave sources for the space-based GW interferometers such as LISA, Taiji, and Tianqin \citep{2008MNRAS.390..192S,LISA_2017,2019MNRAS.486.4044B,2019MNRAS.486.2336D,Tianqin_2016}
However, we emphasize that the massive seed forming halos in those scenarios do not necessarily merge into high-$z$ quasar host galaxies.

In this paper, we consider a new scenario of the massive seed formation in biased, over-dense regions with $\ga 5$ mass variance,
where high-$z$ SMBHs are expected to form \citep{HALOMASS_Wyithe2006}. 
In such intrinsically rare patches of the universe, stronger halo clustering increases the frequency of halo mergers and boosts 
the mean intensity of LW radiation background in the regions.
Therefore, the modest conditions required to form massive seeds with $100-10^4~\msun$ will be naturally satisfied there.
We generate merger trees of the progenitor halos that end up a high-$z$ quasar host, based on the extended Press-Schechter formalism,
and quantify the expected mean LW intensity irradiating the main progenitors and the merger heating rate along with the trees.
By taking into account the environmental input,
the thermal and dynamical evolution of a massive gas cloud in the main progenitor halo is calculated in a self-consistent way.

Among previous studies in literature, \cite{Valiante_2016} investigated the origin of SMBHs using semi-analytical models and found massive BHs seeded in the quasar progenitor halos, depending on their environmental effects.
Recently, \cite{Lupi2021} also proposed a similar idea that massive seed BH formation would be much more efficient in
a biased halo merger tree based on dark matter (DM) only N-body simulation.
They found that in an overdense region, a large number of atomic-cooling halos experience successive merger heating that
counteracts radiative cooling via H$_2$ lines and potentially promote massive seed formation.
However, most of the halos in their samples do not end up in the most massive DM halo that is supposed to be a high-$z$ quasar host.
Instead, we study the statistical properties of the progenitor halos of a high-$z$ quasar host by generating merger trees. 
Moreover, we explicitly follow the evolution of gas clouds in the main progenitors, taking into account merger heating, radiative cooling,
and chemical reaction networks.
Thus, the two studies are complementary.

This paper is organized as follows. In \S\ref{sec:method}, we summarize our construction of merger histories of a quasar host,
the calculation of environmental LW intensity for individual halos, 
and subsequent gas evolution following the underlying halo mass growth.
In \S\ref{sec:result}, we discuss the results of LW intensity, the fraction of promising heavy seed formation sites,
and the distribution of accretion rate realized.
In \S\ref{sec:metal}, we quantify the critical metallicity that affects thermal evolution of gas and the efficiency of metal enrichment,
and discuss caveats of our model.
In \S\ref{sec:discussion}, we show the mass distribution of seed BHs formed in the high-$z$ quasar progenitors.
Finally, in \S\ref{sec:summary}, we summarize the main conclusions of this paper.
\vspace{5mm}

\section{methodology}
\label{sec:method}

In order to investigate the evolution of luminous quasar progenitors that form in rare, overdense regions in the universe at redshift $z\sim6$, we construct the merger history of DM halos up to $z=50$, and model the evolution of the gas properties within the DM halos along each merger tree. 
The processes we model consist of three parts:
(1) We first construct the hierarchical merger history of a quasar host halo using the Monte Carlo merger tree algorithm.
For a $10^9\,\msun$ SMBH powering the luminous quasar at $z\sim6$, the halo mass is estimated to be $M_{\rm h}\sim10^{12}~\msun$ by comparing the growth rate of quasar density indicated from observations with that predicted by the Press-Schechter formalism \citep{HALOMASS_Wyithe2006}. We therefore focus our analysis on halos that grow to $M_{\rm h}=10^{12}~\msun$ at $z=6$.
(2) For a given merger tree, we calculate the LW radiation background
produced by the surrounding star-forming galaxies at each redshift, in order to model the radiative impact on the gas within the halo.
(3) The evolution of the gas in the parent halo of each tree is studied by taking into account the injection of thermal and kinetic energy due to violent merger events, as well as LW irradiation calculated in step (2) that dissociates the gas coolants.
In the following subsections, we describe in detail the three key ingredients.
Throughout the paper, we adopt cosmological parameters estimated by Planck assuming a $\Lambda$CDM universe
\citep{Planck15}, i.e., $\Omega_{\mathrm{m}}=0.307,~\Omega_{\Lambda}=0.693,~
\Omega_{\mathrm{b}}=0.0486,~H_0=67.7 \mathrm{~km} \mathrm{~s}^{-1} \mathrm{Mpc}^{-1}$.

\vspace{2mm}
\subsection{Merger histories of progenitors}
We construct DM merger trees based on the extended Press-Schechter formalism \citep{Press_Schechter1974,Lacey_Cole1993,2000MNRAS.319..168C}
using the {\tt GALFORM} semi-analytic algorithm summarized in \cite{Parkinson2008}.
Our sample consists of $10^4$ merger tree realizations for the DM halos that end up as high-$z$ quasar hosts
with $M_{\rm h}=10^{12}~\msun$ at $z=6$.
For each tree, we adopt a minimum DM halo mass of $M_{\rm h,min}=10^5~\msun$. Halos smaller than this threshold do not 
significantly impact the gas evolution, because the critical virial temperature 
above which gas collapse can be induced by coolant $\HH$ is $\sim 10^3 ~\K$ \citep[see ][]{Haiman1996,Tegmark1997},
corresponding to halo mass higher than $M_{\rm h, min}$ (see also Fig.~\ref{fig:all_trees}).
Reflecting the rarity of quasar host galaxies, the progenitor halos form in highly biased regions with
$\ga 5$ mass variance \citep{Mo_White2002}.
Note that the fraction of all matter in such rare halos is $\la 10^{-7}$.

\vspace{2mm}
\subsection{Lyman-Werner background intensity}
\label{sec:jlw_method}
Due to the photo-dissociation of H$_2$ exposed to LW radiation, we also consider the local LW intensity 
$\jlw$ (at $h\nu=12.4 \rm{eV}$, hereafter in units of $ \rm 10^{-21}erg~s^{-1}~cm^{-2}~Hz^{-1}~sr^{-1}$) in order to follow the gas evolution 
in a given progenitor halo. Along each merger tree, we calculate the cumulative $\jlw$ from neighboring star-forming galaxies (hereafter source halos).
Based on the model developed by \cite{DFM2014}, the basic equations and assumptions we adopt are summarized as below.

We consider a DM halo with mass $M_{\rm h}$ (gas + DM) at redshift $z$, which is supposed to be the main progenitor in a merger tree. The average number of source halos (within mass range $m\pm dm/2$) that populate a surrounding spherical shell 
(at a physical distance $r$ with thickness $dr$) is calculated by
\begin{align}
  \frac{d^2\mathcal{N}(m, r)}{dmdr} dmdr& =  4 \pi r^2 dr (1+z)^3 ~\frac{\mathrm{d} n_{\mathrm{ST}}(m, z)}{\mathrm{d} m}\nonumber\\
  &~~ \times [1+\xi(M_{\mathrm{h}}, m, z, r)]\mathrm{~d} m,
  \label{eq:number}
\end{align}
where $\mathrm{d} n_{\mathrm{ST}} / \mathrm{d} m$ is the mass function of source halos \citep{Sheth_Mo_Tormen2001}, 
and $\rm \xi$ denotes the non-linear bias function \citep{Iliev2003}, which gives the deviation 
(from random) probability of finding a halo with mass $m$ at distance $r$ from the main progenitor. 
%
We set the minimum source halo mass to be
$m_{\mathrm{ac}, z} \simeq 6\times10^6\msun \left(T_{\rm vir}/10^4~\K\right)^{3/2} \left[\left(1+z\right)/31\right]^{-3/2} $, 
where the halo virial temperature is 
just above the hydrogen atomic-cooling threshold of $T_{\rm vir}=10^4~\K$, where radiative cooling by Ly$\alpha$ 
emission leads to star formation. 
In our model, we do not consider the production of LW radiation background by star formation activity 
in less-massive DM halos. 
The maximum mass of source halos is determined so that the LW intensity converges towards the higher 
mass bins, namely in terms of averaged flux, contributions from the $m_{\rm max}$ halos vanish 
due to their low abundance.
The value of $m_{\rm max}$ ranges from $\sim 10^6 \msun $ to $\sim 10^{10} \msun$ and is larger at lower $z$.

Following \cite{DFM2014}, we compute the average LW radiation flux that irradiates the target halo.
The time-averaged production rate of LW photons (per unit stellar mass) emitted from a surrounding source galaxy is approximated by
\begin{equation}
    \left\langle Q_{\mathrm{LW}}(t)\right\rangle=Q_{0}\left[1+\left(t_{6} / 4\right)\right]^{-3 / 2} \mathrm{e}^{-t_{6} / 300},
\end{equation}
where $Q_{0}=10^{47} \mathrm{~s}^{-1} \msun^{-1}$ and $t~(=t_{6} ~\mathrm{Myr})$ is the time after a single star burst 
in the star-forming halo.
Thus, the specific LW luminosity from the halo is calculated by
\begin{equation}
L_{\mathrm{LW}}(m_\star, t)=\frac{h\langle \nu\rangle}{\Delta \nu}\left\langle Q_{\mathrm{LW}}(t)\right\rangle 
\fesc \left(\frac{m_{\star}}{\msun}\right),
\label{eq:LW_ave}
\end{equation}
where the mean frequency and frequency width of the LW band ($11.2\leq h\nu/{\rm eV}\leq 13.6$) 
are set to $\langle \nu \rangle=12.4~\mathrm{eV} /h$ and ${\Delta \nu} = 2.4~\mathrm{eV}/h$.
The total stellar mass is calculated by $m_\star =f_\star (\Omega_{\mathrm{b}} / \Omega_{\mathrm{m}})m$, 
assuming the star formation efficiency to be $f_\star=0.05$.
The escape fraction of LW photons from the halo is assumed to be unity ($\fesc =1$). This value tends to be lower for atomic-cooling halos with $m\gtrsim 10^7~\msun$. 
As a reference, \cite{2015MNRAS.454.2441S} calculated the LW escape fraction for a single PopIII star in an atomic-cooling halo with 1D simulations and 
found $\fesc \simeq 0.7$. However, this is considered to be a lower bound because the escape fraction would be higher for 3D calculations through directions with 
lower optical depths, besides a higher SFR is expected in our case (rather than a single massive star).
We estimate the LW luminosity at one free-fall time after the burst of star formation:
$t_{\mathrm{sf}}=\sqrt{3 \pi/ (32 G \Delta_{\rm vir} \bar{\rho})}\simeq 18~{\rm Myr}~[(1+z)/31]^{-3/2}$, 
where $\Delta_{\rm vir}\simeq 18~\pi^2$.
Using Eqs.~(\ref{eq:number})-(\ref{eq:LW_ave}), we obtain the mean LW radiation intensity in the target halo as
\begin{equation}
  \jlw(M_{\mathrm{h}},z)= \int_{m_{\mathrm{ac}, z}}^{m_{\rm max}} \int_{r_{\rm min}}^{r_{\rm max}} \frac{d^2\mathcal{N}(m, r)}{dmdr} 
  \cdot \frac{L_{\mathrm{LW}}}{16\pi^2r^2} ~dmdr,
\end{equation}
where $r_{\rm min}$ and $r_{\rm max}$ are the minimum and maximum distance of the source halo from the target halo.
In the absence of metal pollution, $r_{\rm min}$ can be safely set by adding the virial radii of the target and source halos. However, metal enrichment of the main progenitor is a main obstacle in the formation scenario of massive seed BHs, because efficient metal-line cooling (and possibly dust thermal emission) will likely lead to gas fragmentation during its gravitational collapse and thus suppress massive star formation. Generically, there are two types of enrichment processes: (1) genetic enrichment due to past star formation episodes in the progenitors, and (2) environmental enrichment owing to metal bubbles created by supernova (SN) explosions in nearby galaxies. In our model, we consider the environmental enrichment process by adopting the minimum distance to 
source halos as $r_{\rm min }=\max \{r_{\rm vir}(M_{\mathrm{h}})+r_{\rm vir}(m), r_{\rm s}(m)\}$,
where $r_{\rm s}$ is the size of the metal-polluted region surrounding the source halo 
\begin{align}
r_{\mathrm{s}}(m, t)=\left(\frac{E_{\rm SN} m_{\star}}{m_{\rm 0} \rho_{\rm s}}\right)^{1 / 5} t^{2 / 5},
\label{eq:SN}
\end{align}
where $m_0=100~\msun$ is the stellar mass budget required to form a SN progenitor and $E_{\rm SN}=10^{51}~{\rm erg}$ is the explosion energy of a SN.
The density $\rho_{\rm s}$ of gas surrounding the wind is considered to be $\Delta \bar{\rho}_{\rm b}$,
where $\bar{\rho}_{\rm b}$ is the IGM baryon density, and $\Delta =60$ corresponding to the typical baryonic 
overdensity of halos at their virial radius for a NFW profile \cite{DFM2014}.
Similar to the production of LW radiation, we estimate the size of metal-enriched bubbles at $t_{\rm sf}$.
We note that metal-enrichment through in-situ star formation should be subdominant because intense LW radiation suppresses star formation in low-mass progenitors (see \S\ref{sec:metal}).
%

On the other hand, the maximum distance in the integration is given by $r_{\rm max} = \left(\lambda_{\mathrm{LW}, 1}-\lambda_{\beta}\right) c /\left[\lambda_{\beta} H(z)\right]$, where the $\lambda_{\mathrm{LW}, 1} = 1110 \AA $ and $\lambda_{\beta}$ are wavelengths of the lowest LW energy and Ly$\beta$ line, respectively \citep[see][]{Haiman1997}.
We consider the redshift effect by cosmic expansion, where $ H(z) = H_0\left[ \Omega_{\mathrm{m}}(1+z)^3+\Omega_{\Lambda}\right]^{1/2}$ is the Hubble constant at redshift $z$ and $c$ is the light speed. LW photons emitted at $r > r_{\rm max}$ are redshifted into one of the Lyman series resonances and are converted into low-energy photons before reaching the target halo. The $r_{\rm max}$ is thus set as an absorbing screen, i.e., we exclude the contribution of $\jlw$ from halos located at $r > r_{\rm max}$.

\vspace{2mm}
\subsection{Energy injection through halo mergers}

The main progenitor halo experiences vigorous halo mergers in the high-$z$ universe. Successive merger events, in particular major mergers, inject energy into the gas in the parent halo. 
At early phase, energy loss through radiative cooling is inefficient, i.e., the cooling timescale is comparable or longer than the Hubble timescale. Gas is heated through shock formation at the halo virial radius in an adiabatic manner. Subsequently, the energy is transported into the halo interior, leading to gas virialization with a nearly constant temperature profile ($ T_{\rm gas} \sim T_{\rm vir}$) across all radii \citep{Wise_Abel2007}. Assuming that the virial equilibrium state is reached after a merger event, 

the virial theorem applies to the gas in the post-merger halo, where the internal and kinetic (turbulence) energy 
is balanced with the gravitational energy as
\begin{equation}
  e_{\rm tot} = e_{\rm th}+ e_{\rm k} + \Phi_{ R_{\rm vir}} = \frac{1}{2} \Phi_{ R_{\rm vir}},
  \label{eq:virial}
\end{equation}
where $e_{\rm tot}$, $e_{\rm th}$ and $e_{\rm k}$ are the total, thermal, and kinetic energy per unit mass, 
and $\Phi_{R_{\rm vir}}$ is the gravitational energy at the virial radius.
In this work, we adopt the NFW potential for DM halos given by
\begin{equation}
    \Phi_{R_{\rm vir}}=-\frac{2k_{\rm B}T_{\rm vir}}{\mu m_{\rm p}}\cdot \frac{\ln(1+c_{\mathrm{vir}})}{\ln (1+c_{\mathrm{vir}})-c_{\mathrm{vir}}/(1+c_{\mathrm{vir}})},
\end{equation}
where $T_{\rm vir}$ is the halo virial temperature, the concentration parameter of the DM density profile
$c_{\mathrm{vir}} =1.9~(M_{\rm h}/10^7\,\msun)^{-0.13}[(1+z)/31]^{-1}$ \citep{Bullock2001},
$k_{\rm B}$ is the Boltzmann constant, $\mu=1.22$ is the mean molecular weight, and $m_{\rm p}$ is the proton mass.
Therefore, the total energy change owing to the halo evolution is given by
\begin{equation}
\Gamma_{\rm mrg} = - \frac{1}{2} \Phi_{R_{\rm vir}} \left(\frac{2}{3} \frac{\dot{M_{\mathrm{h}}}}{M_{\mathrm{h}}}- \frac{1}{1+z}\frac{dz}{dt}\right),   
\label{eq:Gamma_mrg}
\end{equation}
where the first term of the right hand side denotes the energy change associated with mass growth and
the second term represents the cosmic expansion effect.
In the generally turbulent virialized gas, the kinetic-to-thermal energy ratio is equal to 1 around the virial radius, and decreases to $1/3$ at the center \citep[see][]{Wise_Abel2007}.
Adopting this branching ratio of the total injected energy, the thermal and kinetic heating rate associated with mergers
are given by $\Gamma_{\rm mrg,th}=3\Gamma_{\rm mrg}/4$ and $\Gamma_{\rm mrg,kin}=\Gamma_{\rm mrg}/4$,
respectively.
Combining Eqs.~(\ref{eq:virial})-(\ref{eq:Gamma_mrg}), the gas temperature follows the halo virial temperature as 
\begin{equation}
\frac{\dot{T}_{\rm gas}}{\dot{T}_{\rm vir}} = \frac{1}{2}\cdot \frac{\ln(1+c_{\mathrm{vir}})}{\ln (1+c_{\mathrm{vir}})-c_{\mathrm{vir}}/(1+c_{\mathrm{vir}})}.
\end{equation}
This ratio is close to unity for a wide range of ($M_{\rm h}$, $z$) halos of interest, e.g., 
$\dot{T}_{\rm gas}/\dot{T}_{\rm vir}\simeq 1.3$ and $0.81$ for $c_{\mathrm{vir}}=2$ and $10$.
Note that our method is different from that adopted in previous papers \citep[e.g.,][]{Yoshida2003, Lupi2021}, where $T_{\rm gas}=T_{\rm vir}$ is imposed. The treatment allows us to precisely calculate the radiative cooling rates and chemical reaction coefficients, 
which sensitively depend on the gas temperature.

\vspace{2mm}
\subsection{Turbulence and baryonic streaming motion}

The kinetic energy injected through mergers is stored in the halo as {\it turbulence}.
During the viliarization process, turbulence plays an important role on massive star formation \citep[e.g.,][]{McKee_Tan2002}.
Namely, turbulence acts as a source of pressure, which stabilizes the gas against its self-gravity and delays the collapse
until the cloud becomes massive enough to overcome the turbulent pressure.
In addition to turbulence, the baryonic streaming motion relative to the DM produced in the epoch of cosmic recombination at $z_{\rm rec}\simeq 1100$ also significantly delays gas collapse and star formation in protogalaxies.
The streaming velocity is found to follow a Maxwell-Boltzmann distribution with the root-mean-square speed of
$\sigma = 30~{\rm km~s}^{-1}$ at $z=z_{\rm rec}$ and decays as 
$\tilde{v}_{\rm bsm}=v_{\rm bsm}(1+z)/(1+z_{\mathrm{rec}})$ \citep{BSM2010}.
We note that the volume fraction of the universe with streaming velocities of $v_{\rm bsm} \geq A\sigma$ 
is estimated as $\simeq 0.4$, $8\times 10^{-3}$, and $5.9\times 10^{-6}$ for $A=1$, $2$, and $3$, respectively.

Considering both the three-dimensional turbulence and coherent baryonic streaming velocity, we approximate the 
effective pressure by kinetic motion of gas as
\begin{align}
    P_{\rm tur} \approx  \frac{1}{3}\rho v_{\rm tur}^2 + \rho \left[\alpha_0 \tilde{v}_{\rm bsm}(z)\right]^2,
\end{align}
where $v_{\rm tur}^2 = 2\int \Gamma_{\rm mrg,kin}dt$ is the kinetic specific energy accumulated through 
successive mergers and the coefficient of $1/3$ is required to estimate the pressure due to 
isotropic turbulence \citep{Chandrasekhar1951a,Chandrasekhar1951b}. 
With pressure support from turbulence, gas collapse is delayed to different extents,
with varying strengths of the streaming motion.
In this work, we adopt $\alpha_0=4.7$ in our fiducial model, in order to match the delay of collapse 
obtained from cosmological simulations \citep{Hirano18}.
The total gas pressure is therefore defined by $P_{\rm tot}=P_{\rm gas}+P_{\rm tur}$.

\vspace{2mm}
\subsection{Density evolution}
\label{sec:density_evol}

With the energy injection processes defined above, in this section we describe our model for calculating the density evolution of a gas cloud concentrated in a DM halo that grows through successive merger episodes.
There are three characteristic stages of the evolution: (1) initial adiabatic phase, (2) transition to isothermal gas due to radiative cooling, and (3) gravitationally collapsing phase in a runaway fashion.
We model the gas dynamics in these stages based on a one-zone model \citep[e.g.,][]{Omukai2001}, which is often used to follow the physical quantities at the center of a gravitationally collapsing cloud with a self-similar density profile $\rho_{\mathrm{gas}} \propto r^{-2}$.
However, this profile does not apply to gas in hydrostatic equilibrium before the onset of gravitational collapse. Therefore, we construct a new method to model the three characteristic stages in a physically motivated way.

\subsubsection{Adiabatic Stage}

In the early stage, since the gas density is not high enough for radiative cooling to operate through 
collisionally excited transitions, the gas is adiabatically compressed in the DM halo as the underlying 
DM gravitational potential evolves.
In the DM assembly history through mass accretion, the entropy profile $K(r)$ of the adiabatic gas is 
characterized by a power-law outer profile of $K(r) = K_{\rm vir}(r/R_{\rm vir})^{1.1}$, 
and a constant core with $K_0 \simeq 0.1 K_{\rm vir}$, where $K_{\rm vir} = k_{\rm B} T_{\rm vir} / \left[(\mu m_{\rm p}) \bar{\rho}_{\rm b} ^{2/3}\right]$ 
is the gas entropy at the virial radius \citep{Voit2003,Voit2005}. 
This self-similar entropy profile is also found to be established inside high-$z$ protogalaxies formed 
in DM halos more massive than $3\times 10^6 \msun$ at $z=10$, while the core entropy for less massive halos 
is maintained at the IGM entropy when gas decouples from the cosmic microwave background \citep[CMB; see more details in][]{Visbal2014}.
Motivated by both numerical simulations and galaxy cluster observations, we approximate the entropy profile as
\begin{equation}
K(r) \simeq K_{\rm vir} \left(\dfrac{r}{R_{\rm vir}}\right) + K_0,
\end{equation}
where $K_0={\rm max}(0.1K_{\rm vir}, K_{\rm IGM})$.
Using the entropy profile and the equation of state given by $P_{\rm gas}=K(r) \rho_{\rm gas}^\gamma$, where $\gamma=5/3$,
we calculate the density profile by solving the hydrostatic equation (the so-called Lane-Emden equation) for the cloud embedded in the DM potential:
\begin{equation}
    \frac{1}{r^2} \frac{d}{dr}\left[ \frac{r^2}{\rho_{\rm gas}} \frac{d (K \rho_{\rm gas} ^\gamma +P_{\rm tur})}{dr}\right] = 
    - 4\pi G \left(\rho_{\rm gas}+\rho_{\rm DM}\right).
\end{equation}

Throughout this paper, we adopt the NFW density profile of dark matter halos of all masses 
characterized by a simple analytical form of 
\begin{equation}
    \rho_{\rm DM}(r)=\rho_{\rm m}(z) \frac{\delta_{0}}{\left(c_{\mathrm{vir}} r / R_{\mathrm{vir}}\right)\left(1+ c_{\mathrm{vir}} r / R_{\mathrm{vir}}\right)^{2}},
\end{equation}
where $ \rho_{\rm {m}}(z)$ is the mean matter density and
\begin{equation}
    \delta_{0}= \frac{200}{3} \frac{c_{\mathrm{vir}}^3}{\ln (1+c_{\mathrm{vir}})-c_{\mathrm{vir}} /(1+c_{\mathrm{vir}}) }
\end{equation}
is the characteristic overdensity within halo virial radius \citep{NFW97}.

We integrate this hydrostatic equation with respect to $\rho_{\rm gas}(r)$ imposing the regularity conditions
at the center, i.e., $\rho_{\rm gas}=\rho_0$ and $d\rho_{\rm gas}/dr=0$ at $r=0$.
Since the solution for adiabatic gas generally has the radius $r_0$ where $\rho_{\rm gas}(r_0)=0$,
we determine the central density $\rho_0$ so that the enclosed gas mass at $r\leq r_0$ satisfies
$M_{\rm gas}=f_{\rm b} M_{\rm h}$, where $f_{\rm b}=\Omega_{\rm b}/\Omega_{\rm m}$ is the baryonic fraction.

\vspace{2mm}
\subsubsection{Isothermal Stage}

As the gas temperature increases due to gravitational compression and merger heating, 
radiative cooling processes begin to operate in the cloud
and the adiabatic assumption no longer applies.
When the radiative cooling timescale is shorter than the heating timescale,
we solve the hydrostatic equation for the density profile assuming an isothermal equation of state:
\begin{equation}
    \frac{1}{r^{2}} \frac{d}{d r}\left[r^{2} c_{\rm eff}^2 
    \frac{d \ln \rho_{\rm gas}}{d r}\right]=
    -4 \pi G \left(\rho_{\rm gas}+\rho_{\rm DM}\right),
\label{eq:LE_iso}
\end{equation}
where $c_{\rm eff} \equiv \sqrt{ c_{\rm s}^2+v_{\rm tur}^2/3+\left(\alpha_0 \tilde{v}_{\rm bsm}\right)^2}$ is the effective sound speed 
developed from the isothermal sound speed $c_{\rm s} \equiv \sqrt{k_{\rm B}T_{\rm gas}/(\mu m_{\rm p})}$.
The solution of the isothermal Lane-Emden equation with the regularity condition does not have the radius 
where the density becomes zero, but connects to the external medium with a density of $\rho_{\rm ext} = f_{\rm b}\rho_{\rm DM}$. The central density is determined so that $\rho_{\rm gas} = \rho_{\rm ext}$ at the virial radius.

From the analogy of the Bonnor-Ebert sphere, the isothermal gas cloud embedded in a DM halo potential 
has a critical mass for the onset of its gravitational collapse.
Practically, for a given $T_{\rm gas}$ and $\rho_{\rm DM}(r)$, we construct the density profile with different 
values of the gas central density $\rho_0$ and thus obtain $\rho_{\rm gas}(R_{\rm vir})$ as a function of $\rho_0$.
Since this function has a local maximum value and the value decreases with increasing halo mass, a hydrostatic equilibrium solution where $\rho_{\rm gas}(R_{\rm vir})=\rho_{\rm ext}$ no longer exists for $M_{\rm h}\geq M_{\rm h,crit}$ (see Appendix \ref{sec:app}). In this case, the gas evolution is described by the free-fall stage below.

\subsubsection{Free-fall Stage}
Once the gas cloud becomes gravitationally unstable, the evolution of the gas density profile is well described 
by the Penston-Larson self-similar solution \citep{Penston1969,Larson1969}, which has a density profile with a flat core 
of the Jeans scale and an envelope with a power-law density distribution $\rho_{\mathrm{gas}}(r) \propto r^{-2}$. 
The central density increases over the free-fall timescale as
\begin{equation}
    \frac{d \rho_{\rm gas}}{dt} = \frac{\rho_{\rm gas}}{t_{\rm ff}},
\end{equation}
where the free-fall timescale is calculated with
\begin{equation}
    t_{\mathrm{ff}} \equiv \sqrt{ \frac{3 \pi}  {32 G \left(\rho_{\rm gas} + \langle \rho_{\rm DM}\rangle  \right)} },
\end{equation}
where $\langle \rho_{\rm DM}\rangle = \rho_{\rm m}(z) \delta_0$ represents the averaged DM density
\footnote{
The squared density of a NFW profile averaged within the characteristic radius of $R_{\rm vir}/c_{\mathrm{vir}}$ is given by 
$\langle\rho^{2}\rangle = \frac{7}{8} \left[\rho_{\rm m}(z)\delta_0\right] ^{2}$,
independent of the concentration factor $c_{\mathrm{vir}}$.
}.

In the collapsing stage, compressional heating by the self-gravitating gas is taken into account
and the rate is given by
\begin{align}
\Gamma_{\rm comp} \equiv \frac{P_{\rm gas}+P_{\rm tur}}{\rho_{\rm gas}^2}\cdot \frac{d \rho_{\rm gas}}{dt}
= \frac{c_{\rm eff}^2}{t_{\rm ff}}.
\end{align}
We note that the compressional heating rate is enhanced by turbulent pressure through the effective sound speed.

\vspace{2mm}
\subsection{Temperature and chemical evolution}

We consider the evolution of thermal and kinetic energy of the gas by solving the two energy equations:
\begin{align}
    \frac{d e_{\rm th}}{dt} & = \Gamma_{\rm mrg,th} + \Gamma_{\rm comp} - \mathcal{L}_{\rm chem}  
    - \mathcal{L}_{\rm rad},
\end{align}
where $\mathcal{L}_{\rm chem}$ is the cooling/heating rate associated with chemical reactions,
and $\mathcal{L}_{\rm rad}$ is the radiative cooling rate (note that all the rates are in units of erg s$^{-1}$ g$^{-1}$).
While the compressional heating rate is included only in the collapse stage, 
the other effects are taken into account to calculate the gas temperature over the three evolutionary stages.
The cooling term includes radiative cooling by H, He, He$^+$, and He$^{++}$ \citep{Glover_Jappsen2007},
 H$_2$ \citep{Glover_Abel2008,2015MNRAS.451.2082G,2015MNRAS.453.2901G}, and cooling/heating associated with chemical reactions.

\begin{figure}
    \begin{center}
    \includegraphics[width=85mm]{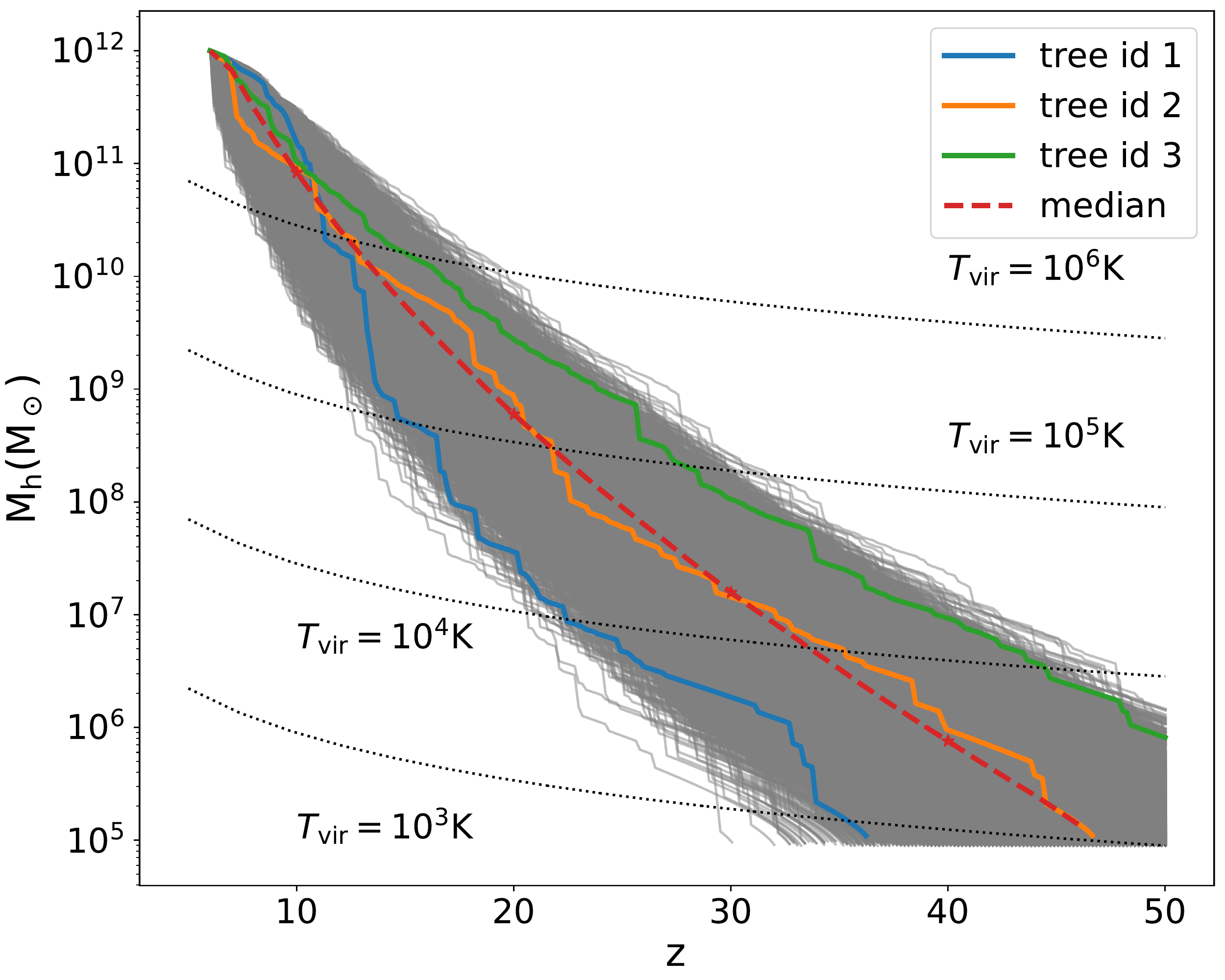}
    \caption{Merger history of the main progenitors of a high-$z$ quasar host galaxy with a DM halo mass
    of $M_{\rm h}=10^{12}~\msun$ at $z=6$.
    For a reference, the median halo mass among all the $10^4$ trees is shown with the red curve.
    Three representative merger trees (in terms of growth speed) are highlighted with the blue, orange, and green curves (tree id = 1, 2, and 3).
    The dotted curves indicate constant virial temperatures, the values of which are denoted by numbers in the figure. 
    }
    \label{fig:all_trees}
    \end{center}
\end{figure}

We solve the chemical reactions of primordial gas among the following 9 species; H, H$_2$, e$^-$, H$^+$, H$^+_2$ , H$^-$, 
He, He$^+$, and He$^{++}$. 
In Table.~\ref{table:reactions}, we show the 35 reaction rate coefficients adopted in this work.
In terms of photodissociation of H$_2$, H$^-$ and H$_2^+$ by external radiation emitted from nearby star-forming galaxies, 
the reaction rate is calculated by assuming the radiation spectral energy distribution (SED) to be a blackbody spectra with $T_{\rm rad}=2\times  10^4 ~\K$.
The SED model approximates more realistic spectra of observed metal-poor star-forming galaxies \citep{Inoue2011}.
The dissociation rates of H$^-$ and H$_2^+$ are calculated by a convolution with the cross section of the $i$-th chemical species ($i=$ H$^-$ and H$_2^+$) as
\begin{equation}
    k_{\mathrm{i}, \mathrm{pd}}=\int_{0}^{\infty} \frac{4 \pi J(\nu)}{h \nu} \sigma_{\mathrm{i}}(\nu) d \nu.
\end{equation}
The cross sections we adopt are from references listed in Table.~\ref{table:reactions}.

\vspace{5mm}
\section{Results}\label{sec:result}

\begin{figure}
    \begin{center}
    \includegraphics[width=85mm]{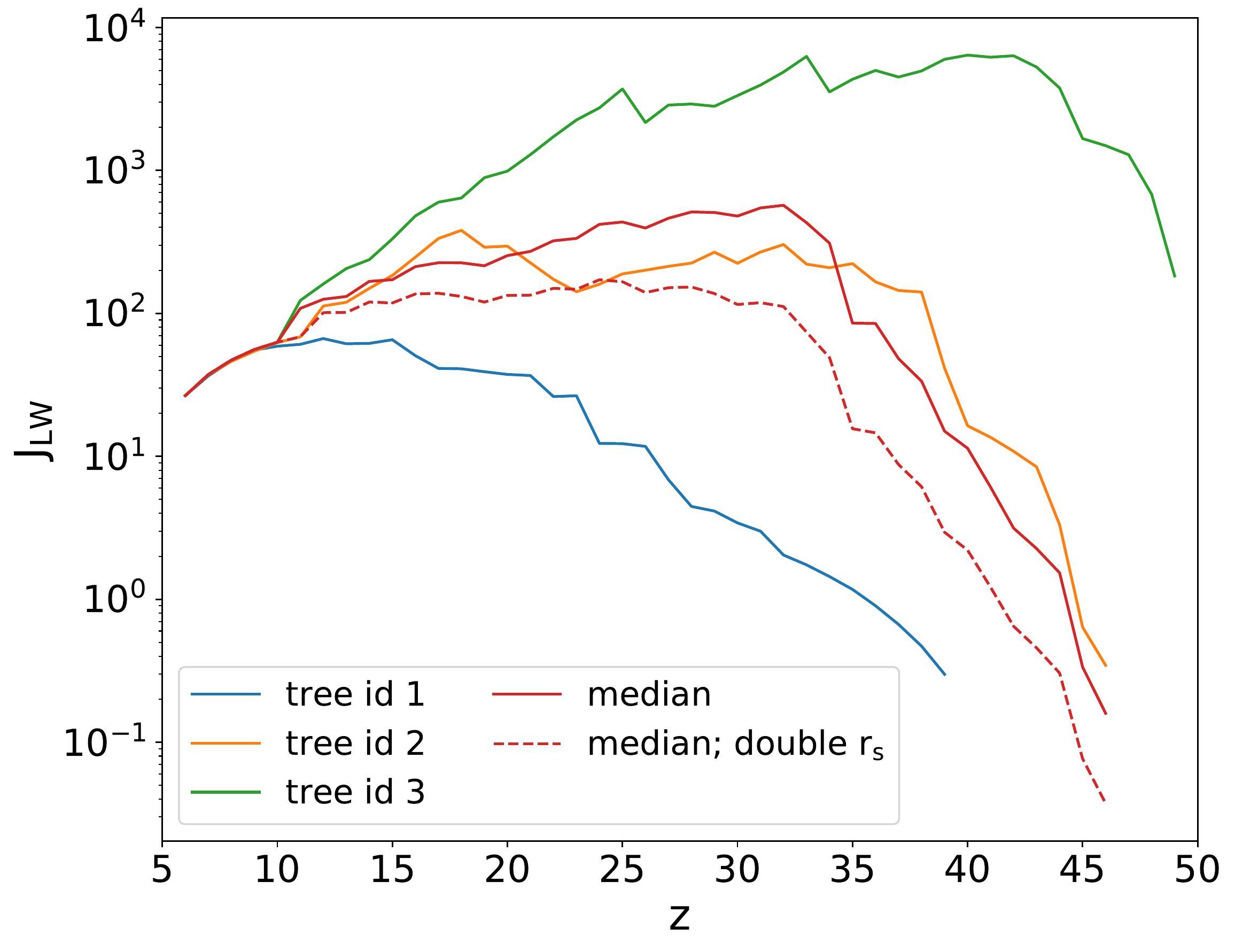}
    \caption{
    Time evolution of LW radiation intensity $\jlw$ (in units of $10^{-21}$ erg s$^{-1}$ cm$^{-2}$ Hz$^{-1}$ sr$^{-1}$) 
    irradiating the quasar progenitors for the four cases shown in Fig.~\ref{fig:all_trees}.
    For the median tree, we show two cases where the metal-bubble size $r_{\rm s}$ is 
    calculated as described in \S\ref{sec:jlw_method} (solid) and the twice of the fiducial value is adopted (dashed).
    }
    \label{fig:Jevol}
    \end{center}
\end{figure}

\begin{figure*}
\begin{center}
\includegraphics[width=170mm]{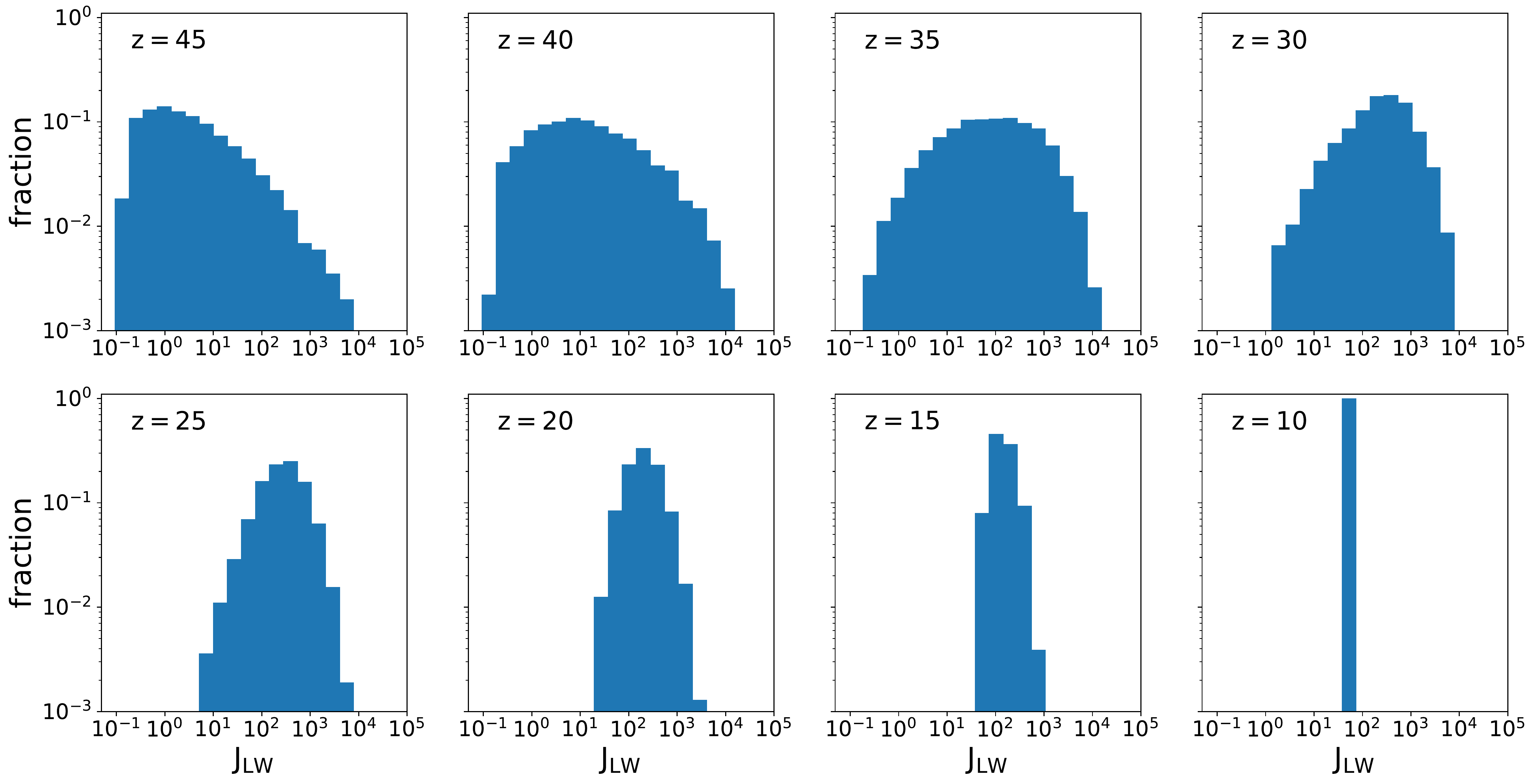}
\caption{
Distributions of the LW background intensity $\jlw$ (in units of $10^{-21}$ erg s$^{-1}$ cm$^{-2}$ Hz$^{-1}$ sr$^{-1}$) 
irradiating the quasar progenitors at different epochs ($10\leq z \leq 45$).
The mean value of $\jlw$ increases from higher redshifts,
has a peak of $\jlw\simeq 450$ at $z\simeq 25$, and decreases toward lower redshifts.
The LW intensity is distributed over a wide range of $10^{-1}\la \jlw \la 10^4$ at higher redshifts, 
while the dispersion of the distribution becomes smaller toward lower redshifts.
}
    \end{center}
    \label{fig:J_histo}
\end{figure*}

\begin{figure*}
    \begin{center}
    \begin{tabular}{cc}
    {\includegraphics[width=83mm]{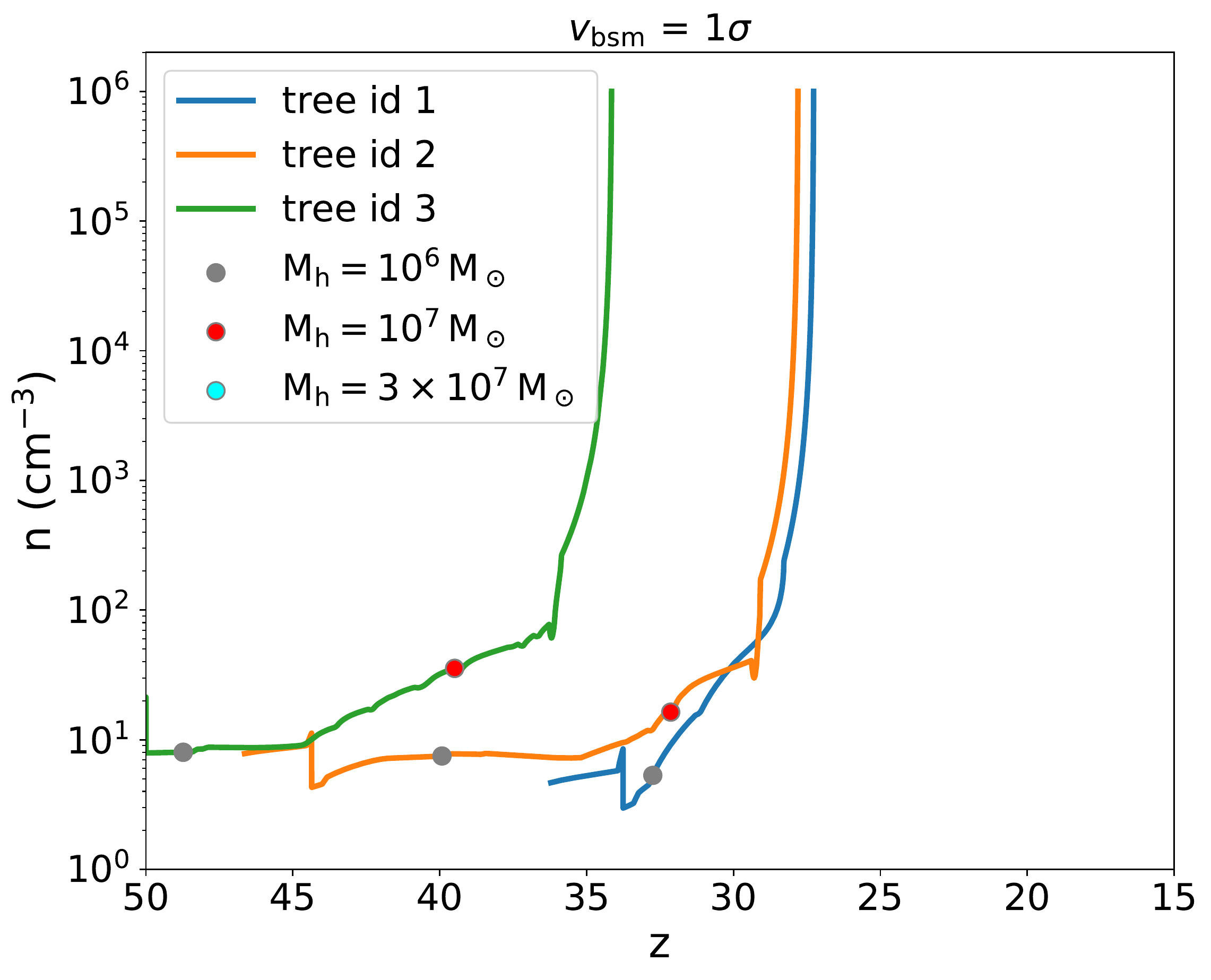}}
    \hspace{5mm}
    {\includegraphics[width=83mm]{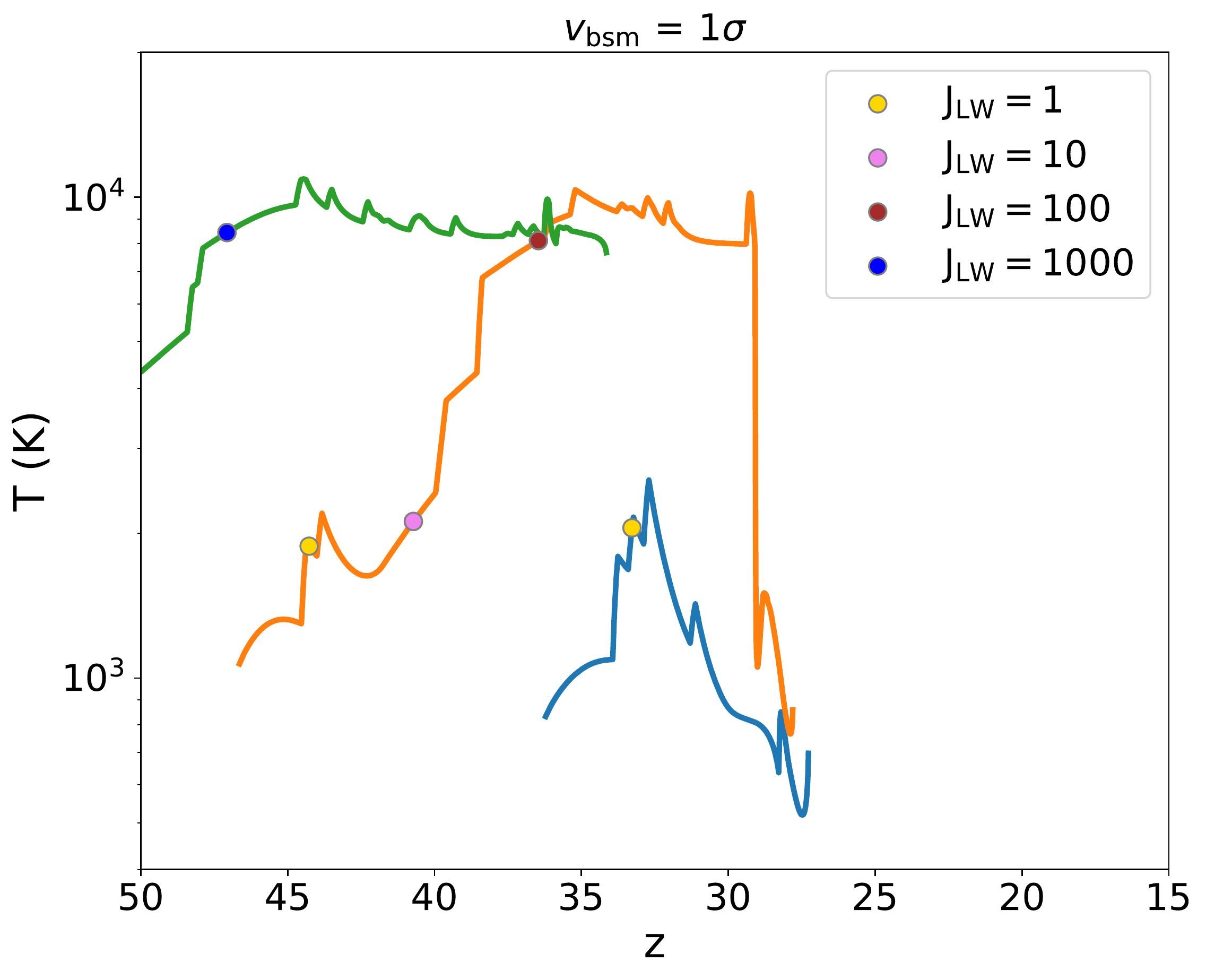}}\\
    {\includegraphics[width=83mm]{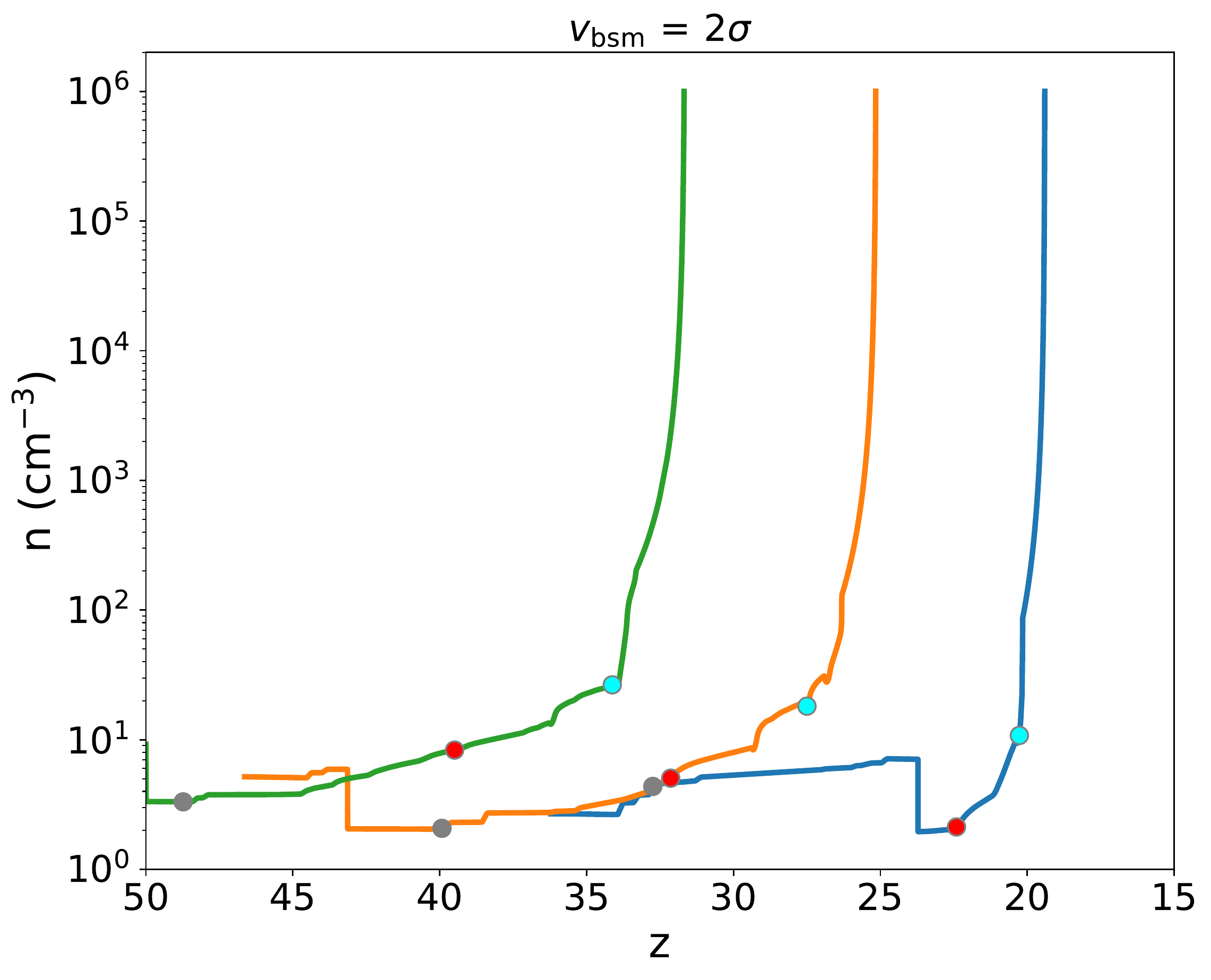}}
    \hspace{5mm}
    {\includegraphics[width=83mm]{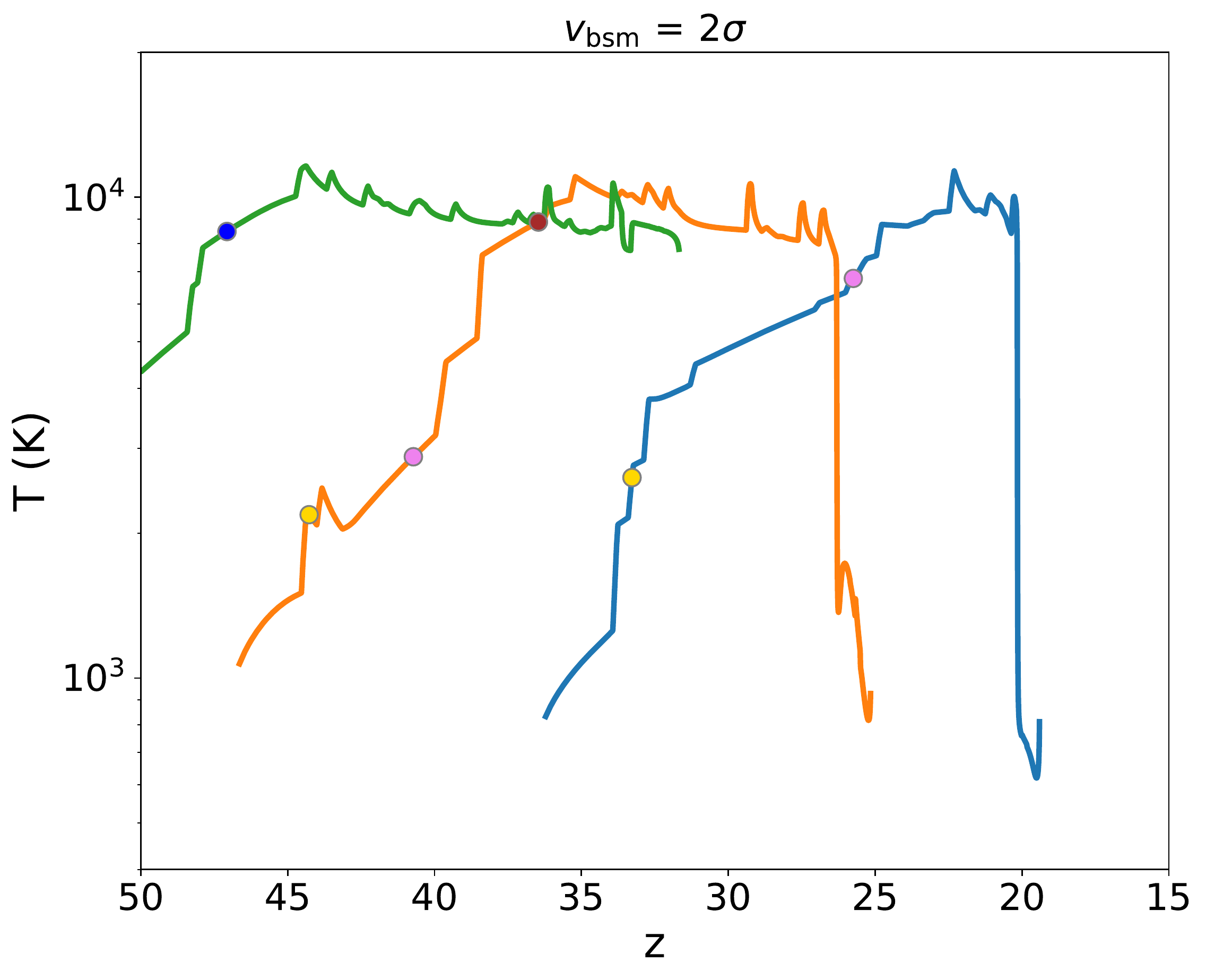}}
    \end{tabular}
    \caption{
    Gas density and temperature evolution along with the three representative halo merger trees
    for the two values of baryonic streaming velocity: $\vbsm = 1\sigma$ (upper panels) and 
    $\vbsm = 2\sigma$ (lower panels).
    The elapsed epochs when the parent halo mass reaches $M_{\rm h}=10^6$, $10^7$ and $3\times10^7\msun$ are marked with dots in the left panels, 
    while those when the LW intensity cross $\jlw = 1$, $10$, $100$ and $1000$ are marked in the right panels.
    When the halo mass grows faster and/or the streaming velocity is higher, gas collapse is significantly delayed due to pressure (thermal + kinetic) 
    support of the gas cloud.
    This effect makes the gas enter the atomic-cooling stage at lower densities (H-H$_2$ and H-H cases) owing to strong LW irradiation
    before the onset of gravitational collapse.
    }
    \label{fig:znzT}
    \end{center}
\end{figure*}

\vspace{2mm}
\subsection{Merger history \& evolution of LW radiation background}
\label{sec:jlw_result}
In Fig.~\ref{fig:all_trees}, we show the evolution of the main progenitors,
i.e., the most massive halos at each epoch, for all the $10^4$ merger trees that grow to $M_{\rm h}=10^{12}~\msun$ at $z=6$.
In such over-dense regions of the universe, the DM halo mass increases via rapid mergers.
The median halo mass (dashed curve) reaches $M_{\rm h} \simeq 8\times10^{10}$, $6\times10^8$, $2\times10^7$,
and $8\times10^5~\msun$ at $z=10$, $20$, $30$, and $40$, respectively,
and the virial temperature exceeds the atomic-cooling threshold of $T_{\rm vir}\simeq 10^4~\K$ at $z\simeq 34$.
Therefore, the gas cloud concentrated in the massive halo
collapses at an epoch earlier 
than when typical first-galaxies would form in atomic-cooling halos
\citep[$M_{\rm h}\simeq 10^7~\msun$ at $z\simeq 10$; see][]{Bromm_Yoshida2011},
which are usually considered to be massive seed forming sites in most previous studies (e.g., \citealt{DFM2014}).
For illustration purposes, we highlight three merger trees: the blue (id 1, a less massive tree), orange (id 2, a tree comparable to the median evolution), and green curve (id 3, a more massive tree). In the following sections, we focus our analysis on these three representative cases.

Following the method laid out in \S~\ref{sec:method}, in Fig.~\ref{fig:Jevol} we present the redshift evolution 
of $\jlw$ for the three representative trees and the median track. 
For all the cases, the LW background intensity gradually increases from higher redshifts, peaks at the intermediate redshifts,
and decreases toward lower redshifts.
This redshift dependence reflects the nature of the non-linear bias function 
which boosts the abundance of halo pairs with comparable masses \citep{2002ApJ...571..585S}.
Namely, when the mass of the main progenitor is close to the atomic-cooling halo mass ($m_{\mathrm{ac},z}\sim 10^7 \msun$), 
a large number of source halos form nearby owing to the halo clustering effect and thus the LW intensity is maximized.
As the main progenitor grows, its mass difference from $m_{\mathrm{ac},z}$ is larger
and thus the clustering effect of atomic-cooling sources becomes weaker so that their spacial distribution is approximated to be uniform (i.e., $\xi\ll1$).
As a result, the LW intensity is dominated by the contribution from a large number of atomic-cooling source halos within the absorbing screen ($r\la r_{\rm max}$) and begins to decline due to the cosmic dilution effect at lower redshifts.
For rapidly growing progenitor halos exceeding $m_{\mathrm{ac},z}$ earlier, the LW intensity quickly rises at higher redshifts and 
the peak values become higher owing to stronger clustering at earlier epochs.
Namely, the peak values of LW intensity in the overdense regions are $\jlw \simeq 60$ (id 1), $\jlw \simeq 400$ (id 2), 
$\jlw \simeq 600$ (median), and $\jlw \simeq 6\times 10^3$ (id 3), which are significantly higher than the level of LW intensity 
irradiating typical atomic-cooling halos that are expected to form massive BH seeds \citep[see][]{Dijkstra2008,Agarwal2012,2013MNRAS.428.1857J}.

In our semi-analytical approach, we model metal pollution of the progenitor halos due to SN explosions that occur in source halos.
Although we treat this effect by replacing the minimum distance between the target and source halos with $r_{\rm s}$, 
there is no information on the time-dependent spatial distributions of DM halos in our framework.
To examine the impact of the model assumptions, in Fig.~\ref{fig:Jevol} we also show the case where the size of the metal-polluted 
bubbles ($r_{\rm s}$) is doubled, the corresponding $t_{\rm sf}$ is comparable to the Hubble time at the redshift,
or equivalent to setting $\Delta=1$ with the fiducial value of $t_{\rm sf}$.
In this case, the LW intensity is overall reduced at higher redshifts,
indicating a significant contribution from nearby source halos with $\ga m_{\mathrm{ac},z}$ to the LW radiation background.
We note that our treatment simply removes the contribution from source halos within distances of $r_{\rm s}$,
but does not address how likely the main progenitor is affected by environmental metal-enrichment.
Our argument nevertheless provides a conservative estimate of $\jlw$ {\it if the efficiency of environmental metal-enrichment is low}.
As discussed in \S\ref{sec:metal}, the efficiency should 
be negligibly low because metal-polluted bubbles rarely penetrate the interior of the target halo \citep{2018MNRAS.475.4378C}.

In Fig.~\ref{fig:J_histo}, we present the histograms of the LW background intensity that irradiates the main progenitor halos for the $10^4$ trees at different redshifts.
For the whole sample of the target halos in highly-biased regions, the histogram resembles a probability distribution function (PDF) of $\jlw$, 
with the bar height in each bin ($\Delta\log \jlw = 0.3$) represents the number fraction of halos irradiated within $\log \jlw \to \log \jlw+\Delta\log\jlw$.
From higher redshifts down to $z\simeq 30$, the mean value of $\jlw$ in the PDF increases owing to a large number of clustered 
source halos with $\ga m_{\mathrm{ac}, z}$ and the $\jlw$ distribution peaks around $\simeq 270$ at $z\simeq 25$. 
Towards lower redshifts, the target halo mass becomes higher than the typical mass of source halos.
Therefore the abundance of sources is hardly boosted by the clustering effect \citep{Iliev2003}.
Moreover, the LW intensity is diluted by the cosmic expansion, lowering the mean value.
While the dispersion of the PDF is larger at higher redshifts, reflecting the diversity of the progenitor mass, 
the PDF peaks at $\jlw \simeq 60$ by $z=10$ when all the $10^4$ trees converge to the high-$z$ quasar host.
We note that our model does not consider LW radiation produced from DM minihalos with $m < m_{\mathrm{ac},z}$,
where $\HH$ is the only coolant to induce star formation.
However, strong LW background radiation in the over-dense region likely suppresses its formation.
Therefore, the histogram shown in Fig.~\ref{fig:J_histo} counts the lower bound of the LW background intensity.

\begin{figure}
    \begin{center}
    \includegraphics[width=85mm]{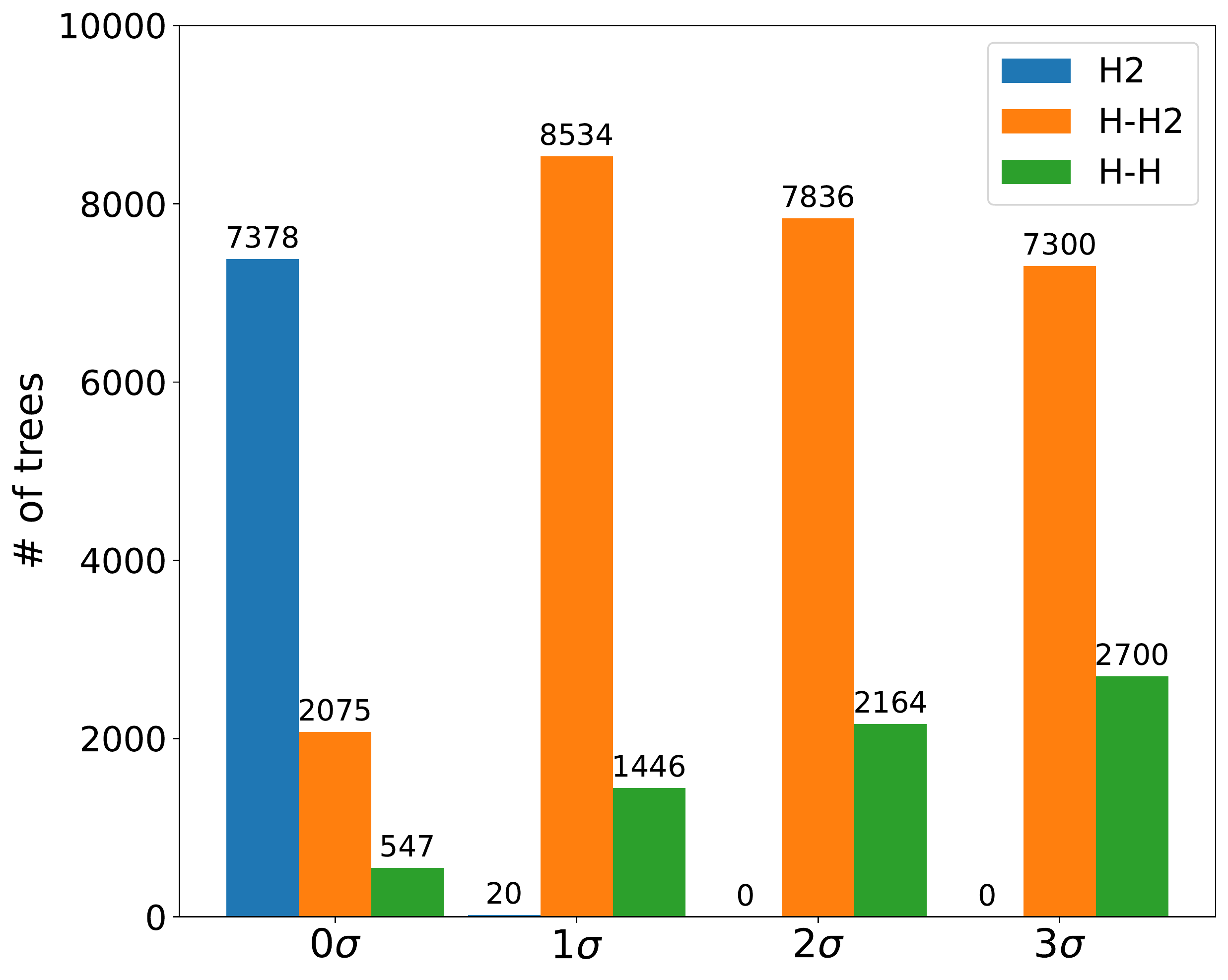}
    \caption{ Census of merger trees which host the three types of gas collapse with different $\vbsm$. 
    The blue, orange and green bars correspond to the representative evolutionary tracks of the same colors 
    in the upper panels of Fig.~\ref{fig:znzT}.
    With increasing $v_{\rm bsm}$, the cases where gas clouds enter the atomic-cooling stage 
    (H-H$_2$ and H-H types) dominate primarily because of the delay of gas collapse that also 
    leads to higher values of LW intensity.    
    }
    \label{fig:count_iso}
    \end{center}
\end{figure}

\begin{figure*}
    \begin{center}
    \includegraphics[width=170mm]{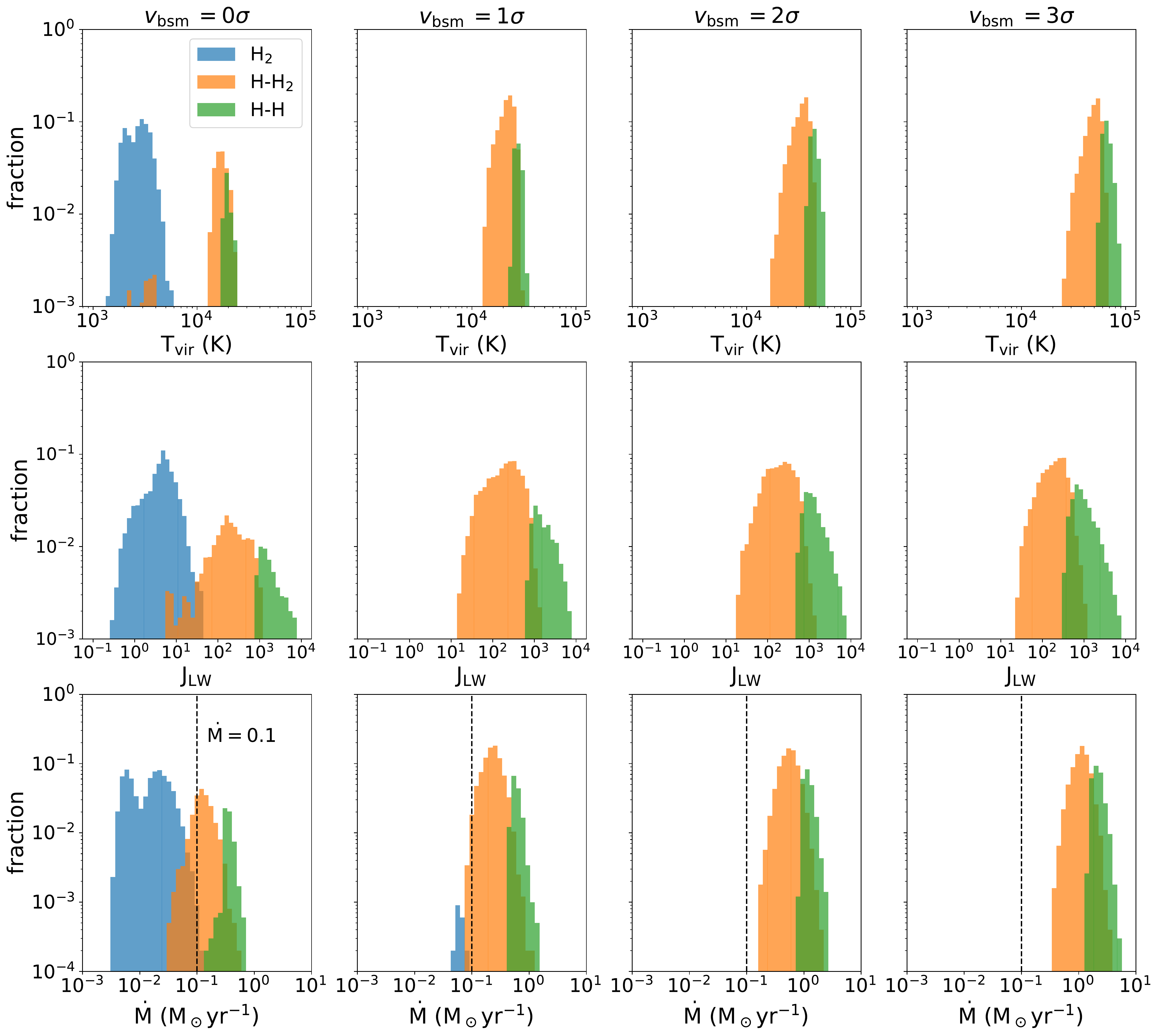}
    \caption{Distributions of the halo virial temperature $T_{\rm vir}$ (upper panels) and LW intensity $\jlw$ (middle panels) 
    measured at the epochs when gas clouds become gravitationally unstable for the cases with different $\vbsm$ values.
    The lower panels show the mass accretion rate of $\Mdot \equiv c_{\rm eff}^3/G$ measured at the minimum temperature point
    at $n_{\rm gas}>10^3\cc$ in the collapsing stage.
    Overall, with $v_{\rm bsm} \geq 1\sigma$, nearly all the cases enter the atomic-cooling stage in massive halos with $T_{\rm vir}>10^4~\K$
    irradiated by LW radiation with intensity of $\jlw >10$.
    Since the collapsing clouds are massive, high accretion rates become high enough ($\dot{M}\ga 0.1~\msunyr$) to form massive seed BHs.
    }
    \label{fig:histo_Tv_J_Mdot}
    \end{center}
\end{figure*}

\vspace{2mm}
\subsection{Thermal and dynamical evolution of gas clouds in the high-$z$ quasar hosts}
\label{sec:gasevo}

In this section we focus our analysis on the gas properties in the main progenitors along the three representative merger trees. In Fig.~\ref{fig:znzT}, we show the evolution of gas density (left panels) and temperature (right panels) at the central core as a function of redshift. In order to examine the impact of baryonic streaming motion, for each merger tree we assume two different $v_{\rm bsm}$ values, i.e., $v_{\rm bsm}=1\sigma$ (upper panels), and $2\sigma$ (lower panels).
Each curve corresponds to the representative case highlighted in Fig.~\ref{fig:all_trees}.
Along with the three evolutionary tracks, we denote the epochs when the DM halo mass exceeds $M_{\rm h}=10^6~\msun$, $10^7~\msun$, 
and $3\times10^7~\msun$ in the left panels, and when the LW background intensity first crosses $J_{\rm LW}=1$, $10$, $10^2$, and $10^3$ in the right panels.
In the following paragraphs, we first describe the gas properties with $v_{\rm bsm}=1\sigma$, and then discuss the impact of the baryonic streaming motion on gas evolution in cases with $v_{\rm bsm}=2\sigma$.

For the lowest mass case (blue curve, tree id 1), the gas density gradually increases with the halo mass in the early stage ($z>30$),
where the gas cloud is supported by thermal and turbulent pressure against its self-gravity and DM gravitational force.
After the halo mass reaches $\simeq 10^6~\msun$, the cloud becomes gravitationally unstable owing to its low temperature,
and collapses over one free-fall timescale at $z\simeq 28$.
The gas temperature remains at $T\la 10^3~\K$ due to H$_2$ cooling, under a modest level of LW intensity ($\jlw \sim 1$) at $z>35$.
In addition to LW radiation, the gas is heated by four major merger events around $z\simeq 31-34$, but the dynamical heating 
rate does not overcome the H$_2$ cooling rate in this case.

For the intermediate mass case (orange curve, tree id 2), the evolution begins from a redshift higher than in the previous case.
In this case, the gas temperature is substantially higher as a result of the combination of merger heating and 
intense LW irradiation with $\jlw \ga 1$ in the early stage.
As several episodes of halo mergers increase the halo mass to $\sim 10^7~\msun$ by $z\simeq 30$ 
(the corresponding halo virial temperature is $T_{\rm vir}\simeq 10^4~\K$), 
the gas temperature reaches $T\simeq 10^4~\K$, where the atomic cooling via Ly$\alpha$ emission begins to operate.
Although the LW intensity reaches $\jlw \ga 100$ before the cloud gravitationally collapses,
the level of LW intensity is not strong enough to suppress H$_2$ formation in the dense region ($\ga 10^2~\cc$),
where H$_2$ reforms owing to its self-shielding effect.
As a result of efficient H$_2$ cooling, the gas temperature drops down to $T\simeq 10^3~\K$ in the collapsing stage.

For the highest mass case (green curve, tree id 3),
the gas temperature quickly rises to $T\simeq 10^4~\K$ due to frequent mergers.
Owing to the clustering effect of the massive parent halo, 
the LW intensity reaches $\jlw \ga 10^3$ at $z\simeq 47$, prominently higher than those seen in the less massive cases.
Although the H$_2$ self-shielding becomes more effective as the central density increases up to $\ga 10^4~\cc$,
the gas collapses keeping a nearly constant temperature of $T\simeq 8000~\K$.
Inside the dense and warm region, H$_2$ is collisionally dissociated and its radiative cooling does not alter the thermal evolution.

In cases where $v_{\rm bsm}=2\sigma$, the gas property evolution is shown in the lower panels of Fig.~\ref{fig:znzT}.
Overall, the collapse of gas clouds is delayed due to kinetic energy injection to the gas concentrated at the halo center.
When the cloud begins to collapse, the corresponding halo masses reach $M_{\rm h} \simeq (3.5,~4.2,~5.9)\times 10^7~\msun$. 
For comparison, the collapse halo masses are $M_{\rm h}\simeq (0.24,~2.1,~2.2)\times 10^7~\msun$ for $v_{\rm bsm}=1\sigma$. 
The delay effect is more remarkable for the lower-mass cases because the halo circular velocity is lower than the effective sound speed boosted by injection of turbulence and streaming motion.
As the gas collapse proceeds, $\HH$ forms efficiently in the modest $\jlw$ environment, 
and eventually its cooling reduces the gas temperature in the low- and intermediate-mass cases.

\vspace{2mm}
\subsection{The statistical properties of the high-$z$ quasar progenitors}

As noted in \S\ref{sec:gasevo} and Fig.~\ref{fig:znzT}, depending on the main cooling processes inducing star formation, 
the evolutionary tracks of the gas clouds embedded in the main progenitors of high-$z$ quasar hosts are classified into three cases: 
(i) $\HH$ cooling, (ii) initial H Ly$\alpha$ cooling followed by $\HH$ cooling after a short isothermal collapse, 
(iii) H Ly$\alpha$ cooling when temperature is kept above $8000~\K$ by compression along a wide density range.
In Fig.~\ref{fig:count_iso}, we present the number count of merger trees for the three types with different baryonic streaming velocities, 
denoted as (i) H$_2$, (ii) H-H$_2$, and (iii) H-H. 
Without the streaming velocity, 74\% of the trees experience gas collapse via H$_2$ cooling, 
while the rest ($26\%$) form atomically-cooling gas clouds (cases H-H$_2$ and H-H).
With non-zero streaming motion ($v_{\rm bsm}\neq 0$), nearly all cases enter the atomic-cooling stage 
because the halo mass reaches $m_{\rm ac,z}$ via mergers due to the significant delay effect.
As the streaming velocity increases, the gas mass becomes higher at the onset of gravitational collapse, 
and thus the compressional heating rate during the collapse stage is higher owing to the accumulation of kinetic energy.
Therefore, the number of trees where gas isothermally collapses with $T\simeq 8000~\K$ (case H-H) increases monotonically 
from $14\%$ to $27\%$ with increasing streaming velocity from $v_{\rm bsm}=1\sigma$ to $3\sigma$.

In Fig.~\ref{fig:histo_Tv_J_Mdot}, we show the distributions of the halo virial temperature (upper panels) 
and LW background intensity (middle panels) for the three types of gas collapse. 
For each case, the values of $T_{\rm vir}$ and $\jlw$ are measured at the epoch when the gas cloud first enters its unstable stage. 
In contrast to cases with $v_{\rm bsm}=0$, where gas collapse is led by H$_2$ cooling in less massive halos with $T_{\rm vir}\sim 10^{3-4}~\K$,
the streaming velocity delays the cloud collapse until after the halo grows across the atomic cooling threshold of $T_{\rm vir}\ga 10^4~\K$.
The virial temperature for the H-H cases is generally higher than that for the H-H$_2$ cases and the mean value of $T_{\rm vir}$ 
for each case increases with larger streaming velocity. 
This trend is more clearly shown in the distributions of $\jlw$,
namely the mean LW background intensity for the H-H cases is $\langle \jlw \rangle \ga 10^3$, 
which is $\simeq 10$ times higher than that for the H-H$_2$ cases. 
The higher value of $\jlw$ is mainly caused by the delay of gas collapse 
until the halo mass becomes massive enough to be exposed by a larger number of LW source halos.
In addition, compressional heating in collapsing clouds is stronger with larger $v_{\rm bsm}$ 
and the minimum LW intensity required to keep isothermal collapse is extended to lower values.

In the main progenitors of high-$z$ quasar hosts, massive gas clouds form owing to the significant delay effect of cloud collapse
by rapid halo mergers and intense LW irradiation from nearby star-forming galaxies.
The mass accretion rate onto the central region of a gravitationally collapsing cloud is approximated as 
$\dot{M} \simeq M_{\rm gas}/t_{\rm ff}$, where $M_{\rm gas}$ and $t_{\rm ff}$ are the gas mass and free-fall timescale 
at the onset of gravitational collapse.
Since the cloud is supported by thermal and kinetic energy of the gas, the accretion rate can be written as $\simeq c_{\rm eff}^3/G$
\citep[][etc.]{Larson1969,Penston1969}, which depends only on the gas thermal and kinetic temperature (see below Eq.~\ref{eq:LE_iso}). 
In the lower panels of Fig.~\ref{fig:histo_Tv_J_Mdot}, we show the distributions of $\dot{M} \equiv c_{\rm eff}^3/G$, for which we adopt the minimum temperature value 
in the cloud collapse stage at $n\ga 10^3~\cc$.
The accretion rate is broadly distributed over $\dot{M} \simeq 3\times10^{-3}-5~\msunyr$.
The vertical line in the bottom panels indicates a reference value of $0.1~\msunyr$,
above which the outer envelope of an accreting protostar is bloated due to rapid heat injection through mass accretion
and the emission of stellar ionizing photons is strongly suppressed.
For $v_{\rm bsm} = 1\sigma$, the majority of the H-H$_2$ cases yield $\dot{M}\ga 0.1~\msunyr$. 
With $v_{\rm bsm} > 1\sigma$, all the cases have sufficiently high accretion rates exceeding the reference value
(see more discusssion in \S~\ref{sec:discussion}).

\vspace{5mm}
\section{Effects of Metal enrichment}
\label{sec:metal}

\subsection{Critical Metallicity}

\begin{figure}
    \begin{center}
    \includegraphics[width=85mm]{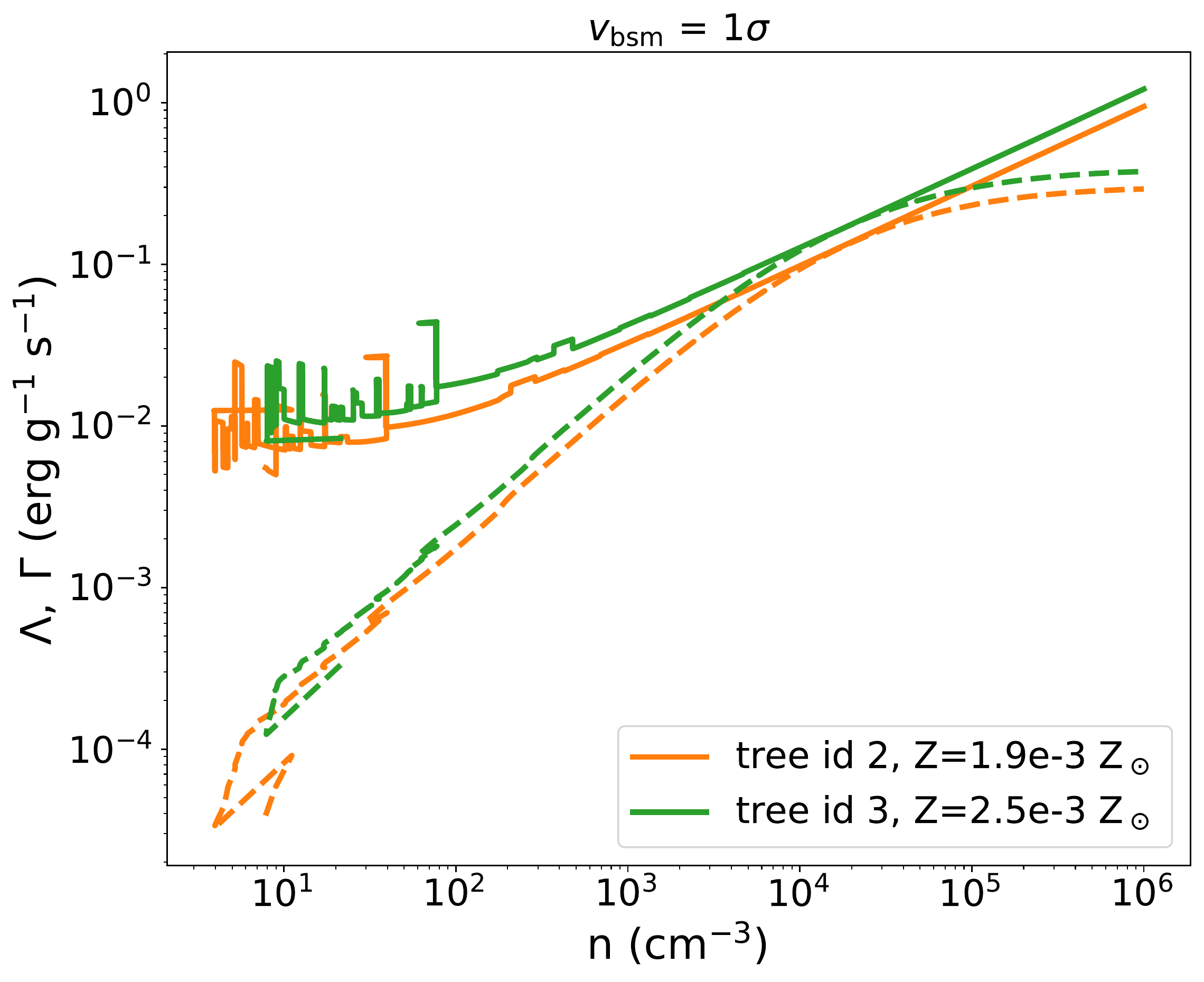}
    \caption{
       Evolution of the heating rate (solid) and metal fine-structure line cooling rate (dashed) with gas density
       for the two representative trees (id 2 and 3) with $\vbsm=1\sigma$.
       The cooling rate consists of C$_{\rm II}$ and O$_{\rm I}$ fine-structure line emission, and 
       the heating rate includes the effect of turbulence and halo mergers.
       To quantify the critical metallicity for which metal-line cooling dominates heating during the gas collapse,
       we turn off the H$_2$ cooling rate.
       The critical metallicity is found to be $Z_{\rm crit}\simeq 1.9\times 10^{-3}~Z_\odot$ and $2.5\times 10^{-3}~Z_\odot$ for the tree 2 and 3, respectively.
    }
    \label{fig:metal_rc}
    \end{center}
\end{figure}

Metal enrichment is considered to be a major obstacle in forming massive BH seeds through star formation 
because efficient radiative cooling via metal fine-structure lines will induce gas fragmentation and  
suppress the formation of masive stars.
In order to quantify the critical metallicity, 
we calculate the cooling rate by C$_{\rm II}$ and O$_{\rm I}$,
assuming that the number fractions of carbon and oxygen nuclei in the gas phase with respect to hydrogen nuclei 
are $x_{\rm C, gas} = 0.927\times10^{-4} (Z/Z_\odot)$ and $x_{\rm O, gas} = 3.568\times10^{-4} (Z/Z_\odot)$ \citep{1994ApJ...421..615P}, 
and all the carbon and oxygen are in the form of C$_{\rm II}$ and O$_{\rm I}$, respectively.
This treatment is justified for warm gas with $T\simeq 8000~\K$ \citep{2008ApJ...686..801O}.

In Fig.~\ref{fig:metal_rc}, we present the metal-line cooling rate (dashed) and heating rate associated with mergers and gravitational compression (solid) 
as a function of the density of gas embedded in the two representative progenitor halos (tree id 2 and 3) with $\vbsm = 1\sigma$.
In order to examine the cooling effect by metal lines against heating, the H$_2$ cooling is turned off, and metal-line cooling is calculated but not included in the thermal evolution.
The metallicity for each case is set so that the cooling rate is marginally balanced with the heating rate 
at least once during the collapse phase.
Namely, the critical metallicity is estimated as $Z_{\rm crit}\simeq 1.9\times 10^{-3}~Z_\odot$ (tree id 2) and $2.5\times 10^{-3}~Z_\odot$
(tree id 3), respectively.
These values are higher than the critical metallicity of $Z_{\rm cirt} \sim 3\times 10^{-4}~Z_\odot$ (in the absence of dust)
obtained by \cite{2008ApJ...686..801O}, where the effect of turbulence and merger heating is not included.
Although the critical metallicity depends on the relative abundance of metals produced in SN ejecta,
we use $Z_{\rm crit}= 2\times 10^{-3}~Z_\odot$ as a reference value in the following discussion.

\subsection{Efficiency of Metal Enrichment}

Throughout this paper, we do not consider the genetic pollution process through mergers of metal-rich minihalos,
given that the star forming efficiency is strongly suppressed by intense LW radiation in the overdense region.
However, we note that this treatment is justified only when the ``actual" LW intensity is as high as the average value 
shown in Fig.~\ref{fig:J_histo}. Otherwise, H$_2$ cooling induces star formation in weak LW-radiation pockets.
We do not quantify this effect that reduces the number of the main progenitors where gas is kept pristine.
As a reference, \cite{Lupi2021} found $\sim 30\%$ of the atomic-cooling halos in the overdense region 
to be polluted genetically.
Since some of those polluted halos do not belong to the merger history of the final massive quasar host halo,
more than 70\% of our main-progenitor samples should remain pristine (or sufficiently metal poor).
On the other hand, together with the metal enrichment effect, we also exclude the contribution of LW flux from such lower mass halos,
making our treatment conservative.

Next, we discuss the modeling of environmental pollution led by SN-driven bubbles from nearby star-forming halos.
One important caveat is that the progenitor halo is assumed to be immediately enriched once the bubble front reaches the halo virial radius.
However, the {\it instantaneous} enrichment process considered in many previous studies in literature may not be realistic.
In fact, metals in SN ejecta cannot penetrate into the halo center but pollute the halo superficially
in the outer region with low densities of $\la 10~\cc$ \citep{2017ApJ...844..111C,2018MNRAS.475.4378C},
leaving the gas in the halo interior un-polluted, even for low mass halos.
If more energetic pair-instability SNe occur in nearby source halos, the ejecta with stronger ram pressure deeply penetrate into 
the target halo and induce metal mixing at the shock front \citep{2017ApJ...844..111C}.
To consider this uncertainty, we introduce the metal mixing efficiency $f_{\rm mix}$, which is
the fraction of metals mixed with the interior gas in the target halo 
and is treated as a free parameter below.

Another important quantity is the total amount of metals carried into the target halo through multiple SN-driven bubbles.
Let us consider a source halo $m$ with a distance of $r_{\rm s}$ from the target halo with a size of $r_{\rm vir}(M_{\rm h})$.
The mass of metals produced by multiple SNe in the source halo is given by $m_{\rm met}=N_{\rm sn}m_{\rm ej}$,
where $N_{\rm sn} \simeq m_\star /m_0$ is the number of SNe and $m_{\rm ej}$ is the average mass of metals produced by one SN.
We here adopt $m_{\rm ej}=0.746~\msun$, which corresponds to the metal ejecta mass produced by a $13~\msun$ stellar progenitor 
\citep{2018MNRAS.475.4378C}.
Assuming that a fraction $f_{\rm esc,m}$ of the metals is launched isotropically by the SN bubble, 
the mass of the metals that reach the target halo is given by $f_{\rm esc,m} m_{\rm met} (r_{\rm vir}/r_{\rm s})^2/4$.
Therefore, due to SN bubbles produced from one source halo, the gas metallicity in the target halo increases by 
\begin{align}
\Delta Z &\simeq \frac{m_\star m_{\rm ej}}{f_{b}M_{\rm h}m_0} \cdot \frac{f_{\rm esc,m} f_{\rm mix}}{4}  \left(\frac{r_{\rm vir}}{r_{\rm s}}\right)^2\\
&\simeq 9.3\times 10^{-5}~Z_\odot~ f_{\rm mix} \left(\frac{f_{\rm esc,m}}{0.5}\right)
\left(\frac{m}{M_{\rm h}}\right)
\left(\frac{5r_{\rm vir}}{r_{\rm s}}\right)^2,\nonumber 
\end{align}
where $f_{\rm esc,m}\simeq 0.5$ is motivated by a 3D high-resolution hydrodynamical simulations of SN-driven galactic outflows \citep{Li_2017}.

As discussed in \S~\ref{sec:jlw_result}, the LW intensity peaks when the target halo reaches the atomic-cooling threshold because
(1) source halos with $m_{\rm ac,z}$ are the most abundant population in number and (2) two halos with comparable mass are strongly clustered.
This circumstance will also maximize the efficiency of environmental enrichment.
Assuming $M_{\rm h}= m_{\rm ac,z}$, we estimate the number of source halos with mass of $m\geq m_{\rm ac,z}$ located within 
$r_{\rm s}$ ($\simeq 5r_{\rm vir}$ typically) from the target halo for the three representative trees as $N_{\rm s} \simeq$ 0.4 (tree id 1), 6 (tree id 2), 
and 86 (tree id 3), respectively.
As a result, the gas metallicity in the target halo is calculated as $Z=N_{\rm s} \Delta Z \simeq 9.3\times 10^{-5}~Z_\odot f_{\rm mix}N_{\rm s}$.
Therefore, we obtain the conditions where the environmental enrichment process affects the thermal evolution of gas in the target halo as $Z>Z_{\rm crit}$, or equivalently
\begin{align}
N_{\rm s} > 21.5 f_{\rm mix}^{-1}. 
\end{align}
Since $f_{\rm mix}\leq 1$, the gas evolution in the main progenitor surrounded by $\la 21$ nearby source halos within $r_{\rm s}$
is unlikely to be affected by metal-line cooling.
On the other hand, if the mixing efficiency is as high as $f_{\rm mix}\ga 0.25$,
metal enrichment will play an important role in changing the gas evolution in rapidly growing halos (tree id 3),
reducing the number fraction of H-H collapse cases (see Fig.~\ref{fig:count_iso}).

Additionally, inhomogeneous density distributions inside the source halos and non-steady SFR that form SNe in the earlier stage
change the velocity and shape of expanding bubbles. 
Those effects could result in either an overestimation or underestimation of the bubble size.
To discuss the efficiency of environmental enrichment precisely, we need to further study a variety of situations with 
different physical parameters as well as the metal mixing efficiency $f_{\rm mix}$.
We leave this to future work.

\vspace{2mm}
\subsection{Dynamical evolution of metal enriched gas}
We quantify the critical metallicity and discuss the impact of metal-line cooling on the thermal evolution of gas clouds.
However, dynamical evolution of a collapsing cloud with $Z\ga Z_{\rm crit}$ that composes of a warm outer-envelope ($T\simeq 8000~\K$) and a cool central core 
has not been fully understood; especially, longterm behavior of the mass inflow rate onto the central newly-formed protostar is still uncertain.
Recent cosmological simulations suggest that rapid mass inflows may occur even with metal pollution above the critical metallicity
in atomic-cooling halos \citep{2020OJAp....3E...9R}, but widespread star formation limits the final mass of the central star to $\la 10^4~\msun$ \citep{2020OJAp....3E..15R}.
On the other hand, when the metallicity is lower than the critical value, the collapsing gas cloud fragments only at the central region 
and forms a compact disk, in which a vast majority of the clumps merge with the central protostar via inward migration 
\citep{2014MNRAS.445.1549I,2020MNRAS.494.2851C}.
As a result, the stellar growth is not quenched by metal pollution.
Future work is needed to investigate the star formation in the overdense regions where high-$z$ quasar form
to quantify the impact of metal pollution on the gas dynamics.

\begin{figure*}
    \begin{center}
    \begin{tabular}{cc}
    {\includegraphics[width=83mm]{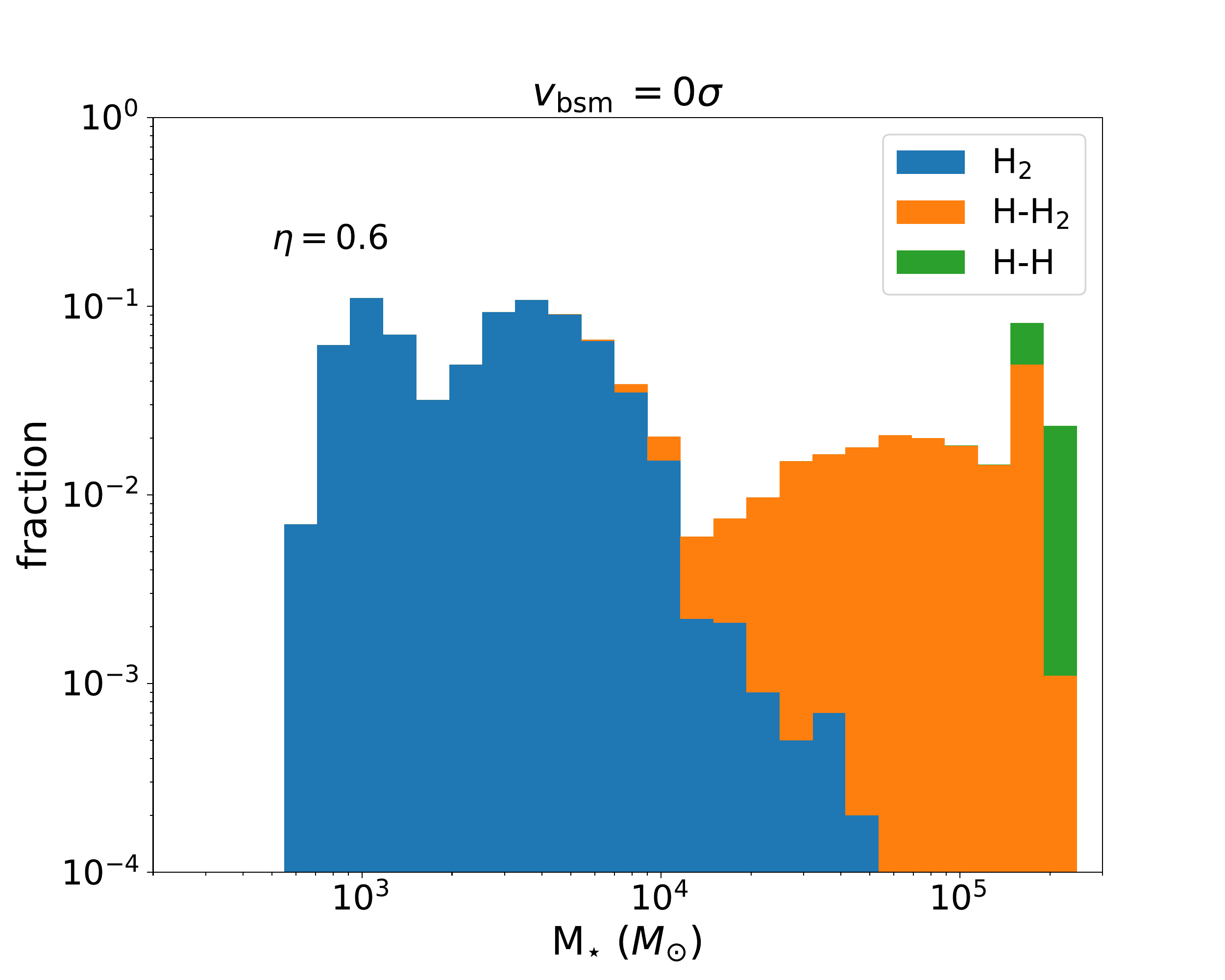}}
    \hspace{3mm}
    {\includegraphics[width=83mm]{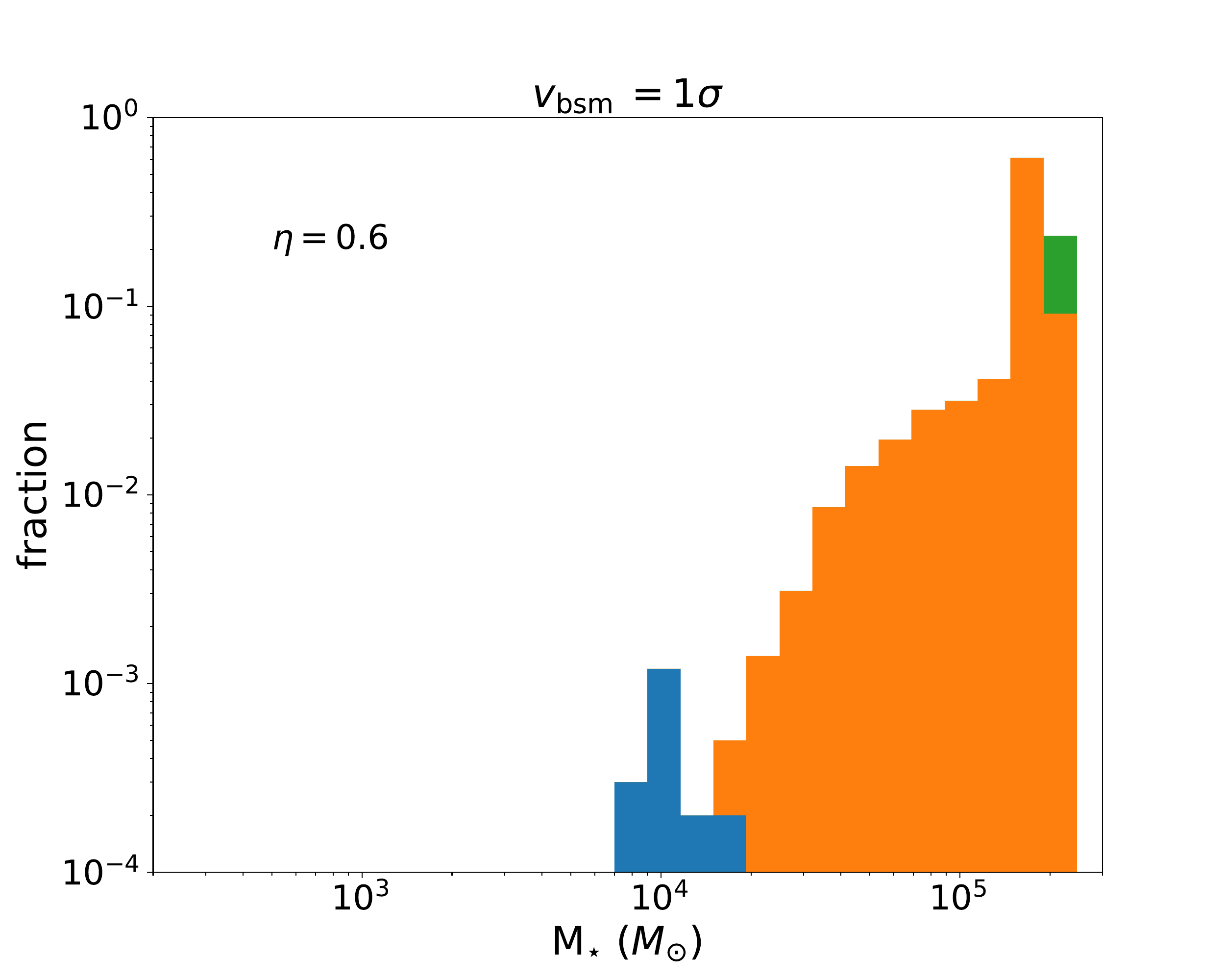}}\\
    {\includegraphics[width=83mm]{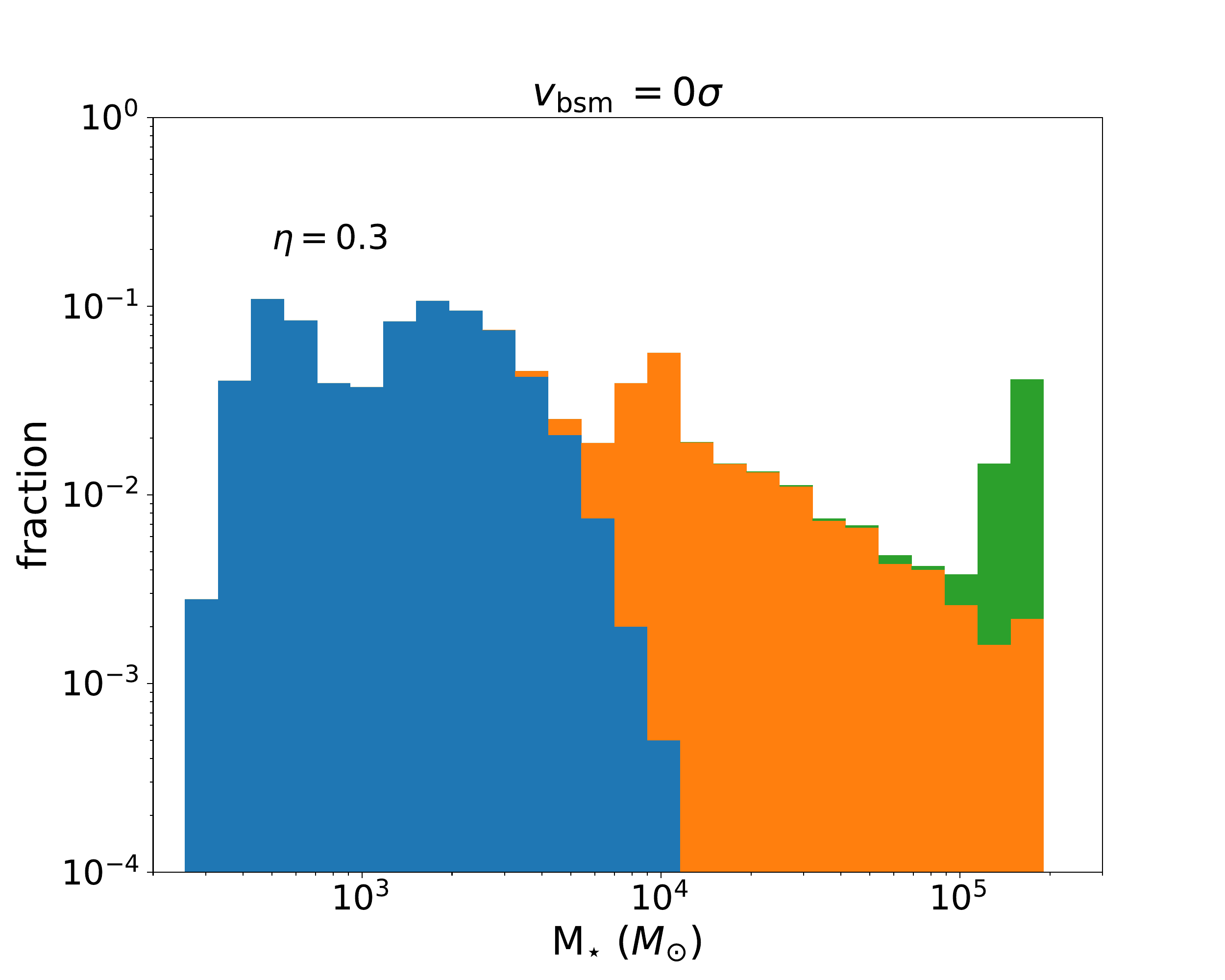}}
    \hspace{3mm}
    {\includegraphics[width=83mm]{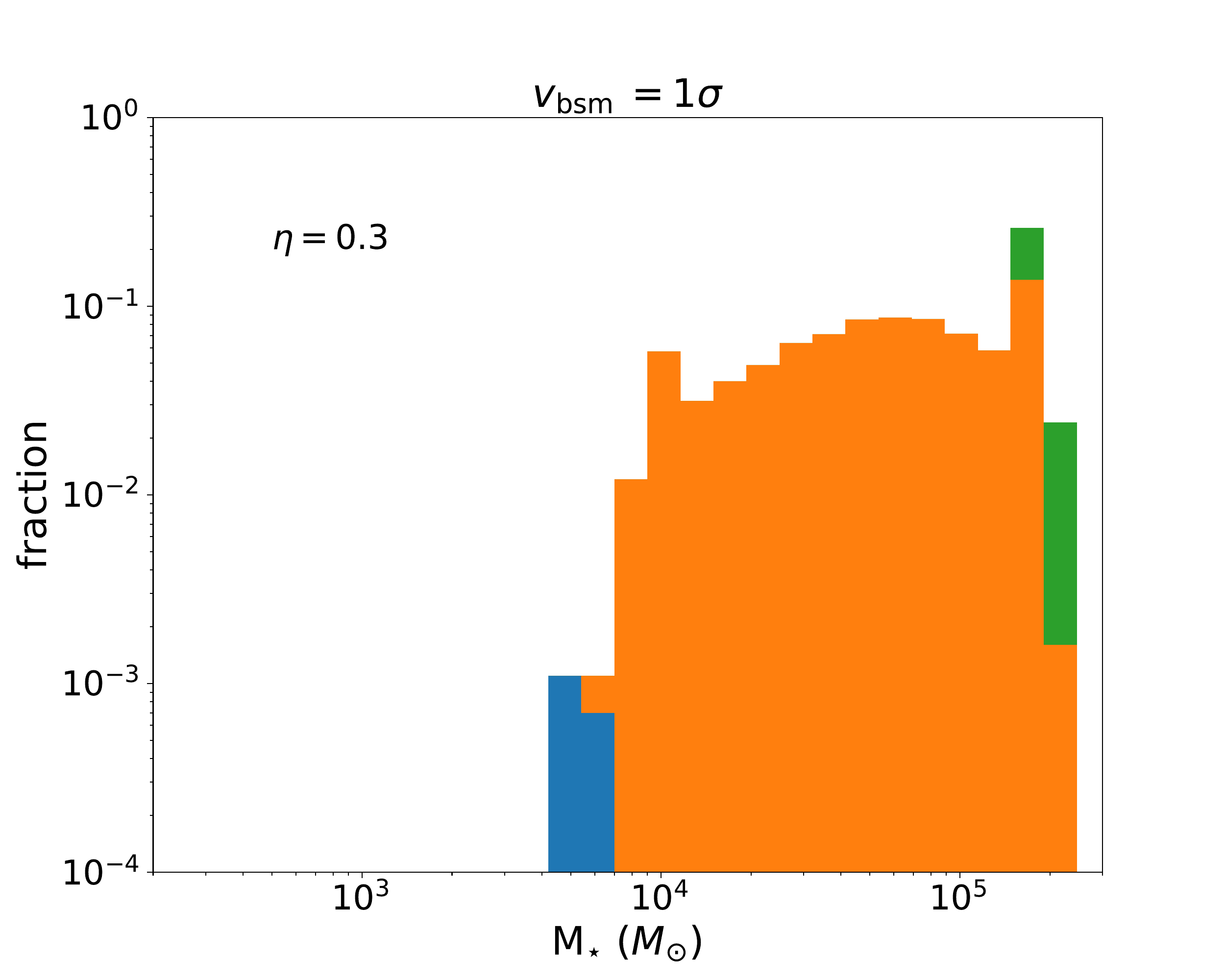}}
    \end{tabular}
    \caption{
    Distributions of the mass of massive stars (equivalent to seed BHs) formed in the quasar progenitors
    with a bin size of $\Delta \log M_{\star}  = 0.1$ for the two cases with $\vbsm=0$ (left panels) and $\vbsm=1\sigma$ (right panels).
    We set the accretion efficiency of $\eta = 0.6$ (and $0.3$) in the upper (lower) panels.
    Without streaming motion, the BH mass is widely distributed from $500~\msun (250~\msun)$ to $\ga 2\times 10^5~\msun$ for $\eta =0.6$ ($0.3$, respectively),
    while for $v_{\rm bsm}=1\sigma$, the lower bound shifts to $7000~\msun (3500~\msun)$.
    }
    \label{fig:MF}
    \end{center}
\end{figure*}

\vspace{5mm}
\section{Discussion}
\label{sec:discussion}

\subsection{Protostellar Mass and BH Mass Distribution}

We apply the obtained mass accretion rate to estimate the final protostellar mass distribution at the end of star formation episodes.
Due to the existing angular momentum at large scales, the rapidly accreted pristine gas settles into a disk,
which becomes gravitationally unstable and thus results in fragmentation and clump formation \citep[e.g.,][]{Oh_Haiman2002}.
Most clumps can migrate inward and merge with the central protostar before forming stars \citep{2014MNRAS.445.1549I},
yielding accretion rate onto the stellar surface through the disk $\Mdot = \eta \dot{M}$,
where $\eta(<1)$ denotes the conversion efficiency from the global accretion rate to that through the accretion disk. 
Hydrodynamical simulations find that mass accretion through the unstable disk proceeds episodically 
and the time-averaged value of the efficiency is $\eta \simeq 0.6$ (\citealt{2016MNRAS.459.1137S}; Toyouchi et al. in prep).

When the time-averaged accretion rate is higher than a critical rate, $\dot{M}_\star \ga \dot{M}_{\rm crit}$,
the accreting star continues to expand its envelope with a lower surface temperature of $T_{\rm eff}\simeq 5000~\K$, 
from which UV radiation hardly emits. 
As a result, stellar radiative feedback does not prevent the central star growing via mass accretion
\citep{Omukai_Palla2001,Hosokawa_2012,Hosokawa2013,2013A&A...558A..59S,Haemmerle2018,Sakurai_2015,Sakurai_2020}.
Since the value of $\dot{M}_{\rm crit}$ ranges from $0.01$ to $0.04~\msunyr$,
depending on the treatment of the stellar evolution calculations and their boundary conditions \citep{Hosokawa2013,Haemmerle2018},
we adopt the highest $\dot{M}_{\rm crit} = 0.04~\msunyr$ as a reference value.
This choice leads to a lower bound of the stellar/BH mass.
With $\dot{M}_\star \ga \dot{M}_{\rm crit}$,
the final stellar mass is determined either by its finite lifetime or by the onset of stellar collapse triggered by
the general-relativistic (GR) instability (\citealt{Shibata_2016}; see a review by \citealt{Woods_2019} and references therein).
The final mass is also affected by fuel supply through mass accretion onto the star.
\cite{2017ApJ...842L...6W} have investigated the final mass of stars accreting at a constant rate over $\simeq 0.01-10~\msunyr$ 
(radiative feedback is neglected), and found that the final mass linearly increases with the accretion rate below $\sim 0.03~\msunyr$
but is saturated around $\sim {\rm a~few}\times 10^5~\msun$ due to the GR instability.
The relation between the critical mass and accretion rate is fitted as 
\begin{equation}
M_{\rm \star,GR} \simeq \left[0.83 \log \left(\frac{\dot{M}_\star}{~\msunyr}\right ) + 2.48 \right] \times 10^5~\msun,
\end{equation}
at $\dot{M}_\star \geq 0.1~\msunyr$, which is used for our analysis.

On the other hand, when the stellar accretion rate is lower than the critical rate, $\dot{M}_\star \la \dot{M}_{\rm crit}$,
the star evolves to the main-sequence stage and begins to emit strong ionizing radiation, quenching the stellar growth.
Here, we simply consider that ionizing radiation from the star heats the disk surface and thus photoevaporation suppresses 
the accretion rate \citep{McKee_Tan_2008,Hosokawa_2011,Tanaka_2013}.
This process becomes important when the ionization front reaches the stellar gravitational influence radius for ionized gas with
a temperature of $2\times 10^4~\K$ defined by
\begin{equation}
R_{\rm inf,\star} \equiv \frac{GM_\star}{c_{\rm s,ion}^2}\simeq 0.17~{\rm pc}\left(\frac{M_\star}{10^4~\msun}\right),
\end{equation}
and the ionized gas breaks out through the neutral infalling gas.
The photoevaporation rate can be expressed as
\begin{equation}
\dot{M}_{\rm pe}\simeq 2.1\times 10^{-2}~\msunyr \left(\frac{\Phi_{\rm ion}}{10^{52}~{\rm s}^{-1}}\right)^{1/2}
\left(\frac{R_{\rm disk}}{0.1~{\rm pc}}\right)^{1/2},
\end{equation}
where $\Phi_{\rm ion}$ is the ionizing photon number flux and $R_{\rm disk}$ is the size of the accretion disk.
The photon flux is approximated as $\Phi_{\rm ion}\simeq 1.6\times 10^{52}~{\rm s}^{-1}(M_\star/10^4~\msun)$ in the range of
$10^3 \la M_\star/\msun \la 10^5$ for main-sequence stars \citep{Johnson_2012}.
We evaluate the mass outflow rate owing to photoevaporation by setting $R_{\rm disk}\simeq R_{\rm inf,\star}$ as
\begin{equation}
\dot{M}_{\rm pe,min}\simeq 3.5\times 10^{-2}~\msunyr \left(\frac{M_\star}{10^4~\msun}\right),
\end{equation}
which gives a lower bound for the rate because the outflow of ionized gas is mainly driven from larger radii (i.e., a lager surface area).
Therefore, equating $\dot{M}_\star = \dot{M}_{\rm pe,min}$, we obtain the feedback-regulated stellar mass as
$M_{\rm \star,fb}\simeq \dot{M}_\star t_{\rm pe}$ or 
\begin{equation}
\dot{M}_{\rm \star,fb}\simeq 2.9\times 10^3~\msun \left(\frac{\dot{M}_\star}{0.01~\msunyr}\right),
\end{equation}
at $\dot{M}_\star \leq \dot{M}_{\rm crit}$, 
where $t_{\rm pe}(\simeq 2.9\times 10^5~{\rm yr})$ is the characteristic 
photoevaporation timescale (note that this expression is valid when the stellar lifetime is longer than $t_{\rm pe}$).
The final mass at the intermediate accretion rate ($\dot{M}_{\rm crit} \leq \dot{M}_\star \leq 0.1~\msunyr$) is estimated by 
performing logarithmic interpolation.

In Fig.~\ref{fig:MF}, we show the mass distribution of massive BH seeds formed in the high-$z$ quasar progenitor halos,
calculated with the method described above (see also the bottom panels in Fig.~\ref{fig:histo_Tv_J_Mdot}).
Note that the number fraction from the different types of gas evolution is stacked at the same mass bin.
Without the streaming motion ($v_{\rm bsm}=0$; left panels), the BH mass is widely distributed from $500~\msun$ ($250~\msun$)
to $\ga 2\times 10^5~\msun$ for $\eta=0.6$ ($0.3$, respectively) with a few peaks corresponding to the virial temperatures of halos
when the BHs form by gas collapse.
Overall, the cases with high accretion rates $\dot{M}_{\rm in}$ (H-H$_2$ and H-H cases) are responsible for high-mass BH 
formation beyond $\sim 10^4~\msun$, while the H$_2$ case with lower accretion rates yields less massive BHs with $<10^4~\msun$.
The number of BH seeds above $2\times 10^5~\msun$ is limited because the GR instability induces direct collapse of accreting supermassive stars.
The shape of the mass distribution at $10^4 \la M_\bullet /\msun \la 10^5$ depend on the accretion efficiency;
namely, the smaller value of $\eta(=0.3)$ yields a distribution skewed toward lower masses.
With non-zero streaming motion ($v_{\rm bsm}=1\sigma$; right panels), the less massive population with $<10^4~\msun$ decreases abruptly
since nearly all the cases experience the atomic-cooling stage and thus the central stars accrete at high rates
without strong radiative feedback.
We note that the BH mass distribution for higher streaming velocities are similar to that for $v_{\rm bsm}=1\sigma$,
but their contribution to the total BH mass distribution is less important because regions with $v_{\rm bsm}\geq 2\sigma$ are rarer.

As discussed in \S\ref{sec:metal}, the number fraction of the cases with highest mass accretion rates (H-H cases)
would be reduced by the effect of line cooling via atomic carbon and oxygen which are produced in nearby source halos through 
SNe and carried into the quasar main progenitor halos with interest.
The level of reduction depends on the metal mixing efficiency in the main progenitor;
namely, the enrichment effect could be neglected if the mixing efficiency is lower than $\sim 20\%$.
Nevertheless, the overall shape of the BH mass distribution still holds.

\subsection{Subsequent BH growth and evolution}

How do those massive seed BHs formed in overdense regions grow to be SMBHs that are observed 
as high-$z$ quasars at $z\simeq 6-7$?
In previous studies in literature, the subsequent growth of their BHs via gas accretion and/or mergers and 
the required conditions have been discussed \citep[e.g.,][]{Tanaka_Haiman_2009,Valiante_2016}. 
Recently, large-scale cosmological simulations have been exploring the evolution of SMBHs and the coevolution 
of their host galaxies including various feedback processes due to SNe and AGN activity with subgrid models.
These simulations have generally found that massive seed BHs formed in protogalaxies hardly grow via gas accretion 
because dense, cold gas is expelled by energetic SN feedback associated with star formation.
However, it is worth noting that most simulations in which SN feedback quenches BH growth have focused 
on ``typical" atomic-cooling halos that will grow to $\sim 10^{10-11}~\msun$ by $z\simeq 6$ \citep[e.g.,][]{Habouzit_2017,Latif_2018,Smith_2018}

On the other hand, as pointed out by \cite{2020ARA&A..58...27I}, the progenitor halos of high-$z$ quasar hosts with 
$M_{\rm h}\simeq 10^{12}~\msun$ at $z\simeq 6$ form 
in rarer regions and have reached $M_{\rm h} \sim 10^8~\msun$
with deeper gravitational potential by the time when star formation takes place ($z\sim 20-35$).
In such massive halos, a large amount of cold gas is supplied to the nuclear region
through filamentary structures of the proto-cosmic web \citep{DiMatteo_2012}, and 
the seed BHs can be fed at high rates significantly exceeding the Eddington limit when the metallicity of inflowing gas 
is as low as $\la 0.01~Z_\odot$ (\citealt{Toyouchi_2021}; see also \citealt{Inayoshi_2016}).
The critical halo mass required for the onset of rapid mass accretion exceeding the Eddington rate is $M_{\rm h}\simeq 10^9~\msun$,
almost independent of redshift.
Most of the quasar progenitor halos of interest can reach this mass threshold after birth of seed BHs
in $\simeq 20-50$ Myr, within which intense star bursts would take place and form protogalaxies.
Exploring the nature of BH growth embedded in such a protogalaxy is left for future investigations.

This process is a possible way to form intermediate massive BH (IMBH) populations.
Observations of IMBHs in the local universe have the potential to constrain high-$z$ BH (seed) formation 
(see the review by \citealt{Greene_2020}).
Furthermore, if those IMBHs form binaries through galaxy mergers and dynamical processes during the cosmic history,
the seed forming channel also provides a significant number of gravitational wave events \citep[e.g.,][]{2018MNRAS.479L..23H,Chon_2018,2020OJAp....3E..15R}, which will be detectable by 
the space-based gravitational wave detectors such as the Laser Interferometer Space Antenna (LISA) \citep{LISA_2017} and Tianqin \citep{Tianqin_2016},
and third-generation terrestrial instruments.

\vspace{5mm}
\section{Summary}
\label{sec:summary}

In this paper, we investigate a new scenario of the formation of heavy BH seeds through collapse of warm gas in massive halos 
that end up in quasar hosts at $z\simeq 6-7$.
In the highly biased, overdense regions of the universe, stronger halo clustering increases the frequency of halo mergers 
and boosts the mean intensity of LW radiation background produced from star-forming galaxies.
Those effects are expected to increase the probability of massive seed formation because the conditions required for 
their formation (intense LW irradiation and violent merger heating) become less stringent than previous considered.
Under such unique environments, we model the thermal and dynamical evolution of massive gas clouds 
along with $10^4$ merger trees of the main progenitors of high-$z$ quasar hosts using the Monte Carlo method.
With those samples, we study the statistical properties of the progenitor halos of high-z quasar hosts and massive seed BHs.
Our major findings can be summarized as follows.

\begin{enumerate}

\item 
In the high-$z$ quasar forming regions, DM halos are likely irradiated by strong LW radiation 
with intensity of $\jlw \simeq 100-10^3$ (in units of $10^{-21}~{\rm erg}~{\rm s}^{-1}~{\rm cm}^{-2}~{\rm Hz}^{-1}$) 
from nearby star-forming galaxies at $z \la 30$ and gas clouds in the halo interiors are heated by successive gaseous halo mergers.
Suppression of H$_2$ cooling via LW irradiation/merger hating as well as injection of gas kinetic energy through halo mergers
prevent gas collapse and delays prior star formation episodes.

\item
Without baryonic streaming motion, 74\% of the trees experience gas collapse led by H$_2$ cooling, 
while the rest (26\%) form atomically-cooling gas clouds that begin to collapse isothermally with $T\simeq 8000~\K$ via Ly$\alpha$ cooling.
With a streaming velocity higher than the root-mean-square value, gas clouds for nearly all $10^4$ realizations of the merger trees 
enter the atomic-cooling stage.

\item
The fraction of trees which host isothermal gas collapse is $14\%$ and increases with streaming velocity, 
while the rest form H$_2$-cooled cores after short isothermal phases.
However, this fraction is reduced by additional cooling via metal fine-structure lines when the collapsing gas could be enriched to 
$Z_{\rm crit}\sim 2\times 10^{-3}~Z_\odot$, requiring efficient metal mixing $f_{\rm mix}\ga 0.25$.
This high probability reflects that high-redshift quasar forming regions likely provide such peculiar environments,
which hardly occur in typical high-redshift star-forming regions.

\item
The mass accretion rate onto a newly-born protostar is distributed over $3\times 10^{-3} - 5~\msunyr$,
a large fraction of which exceeds the critical rate suppressing stellar radiative feedback.
As a result, we expect a distribution of stellar masses (presumably BH masses) ranging from several hundred
to above $10^5~\msun$.

\end{enumerate}

\acknowledgments
We greatly thank Gen Chiaki, Zolt\'an Haiman, Tilman Hartwig, Alessandro Lupi, and Daisuke Toyouchi for constructive discussions.
This work is supported by the National Natural Science Foundation of China (12073003, 12003003, 11721303, 11991052, 11950410493),
the National Key R\&D Program of China (2016YFA0400702), 
and the High-Performance Computing Platform of Peking University.
Y.Q acknowledges support from the China Postdoctoral Science Foundation (2020T130019).


\appendix
\section{The critical conditions for collapse of an isothermal gas cloud}
\label{sec:app}
\begin{figure*}
    \begin{center}
    \begin{tabular}{cc}
    {\includegraphics[width=83mm]{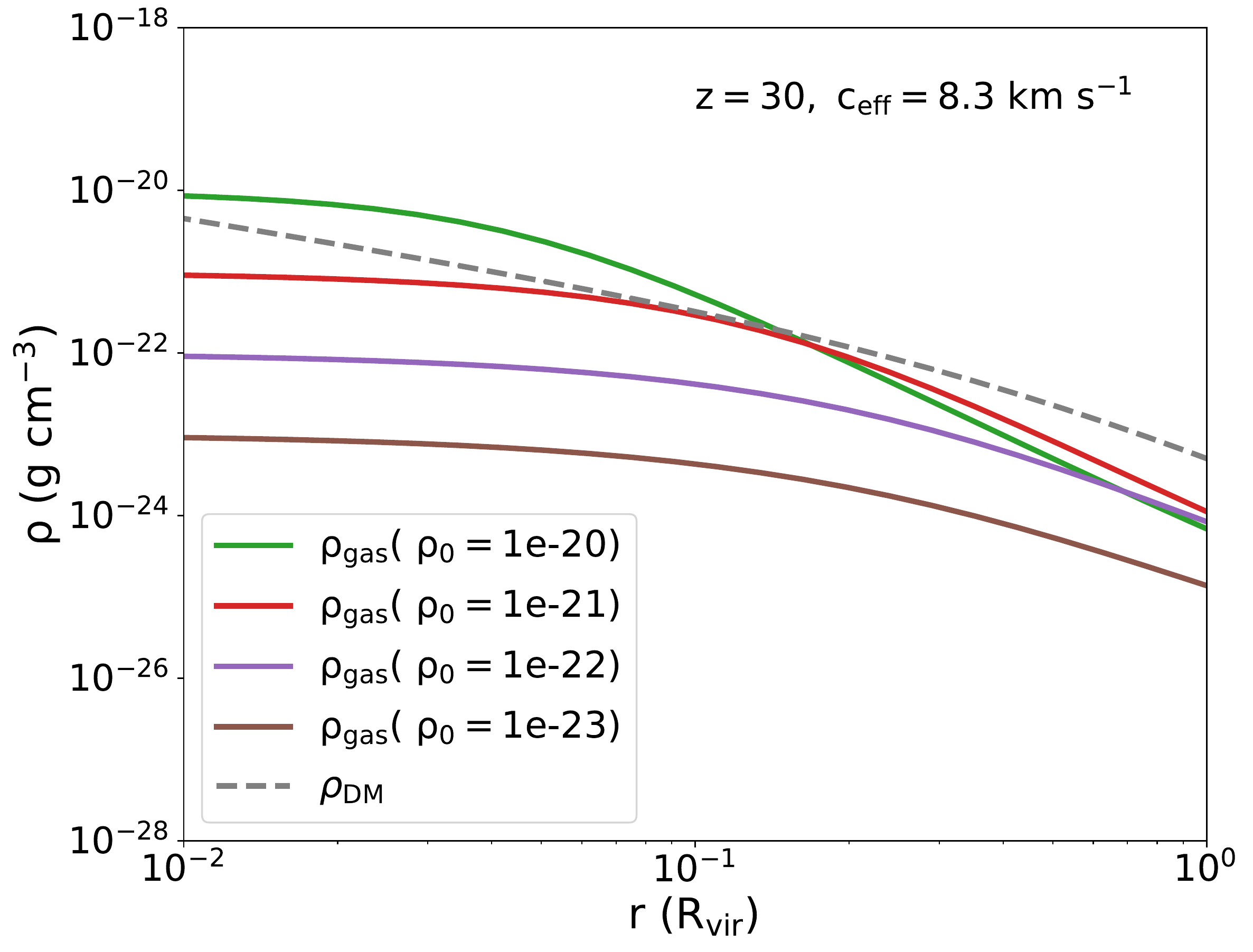}}
    \hspace{5mm}
    {\includegraphics[width=83mm]{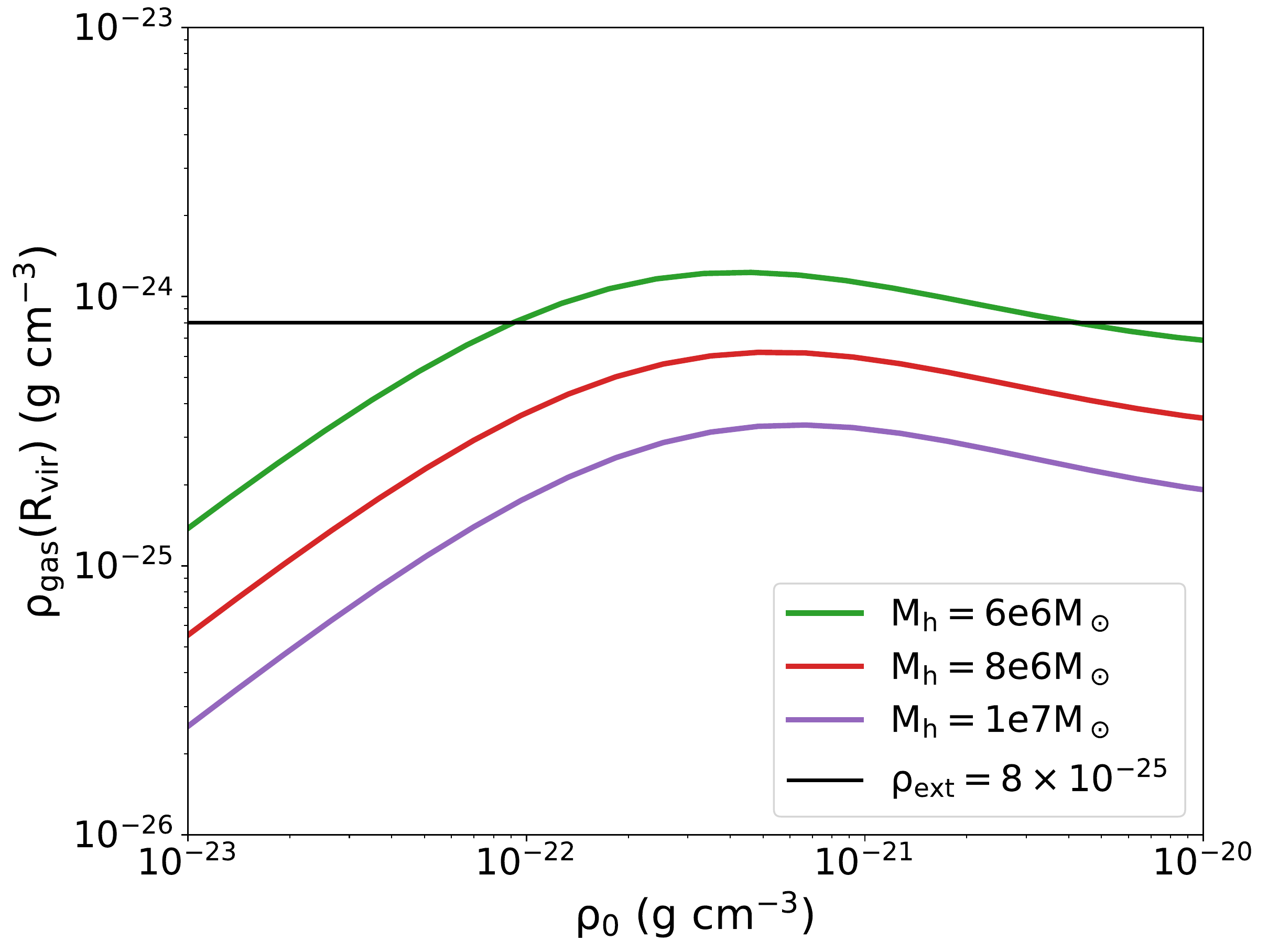}}
    \end{tabular}
    \caption{
    Left panel: gas density profile in a halo with $M_{\mathrm{h}}=6\times10^6~\msun$ at $z=30$, 
    $c_{\rm eff} = 8.3 \mathrm{~km~s^{-1}}$, calculated from $\rho_0 =  10^{-23, -22, -21, -20}\gcc$.
    With increasing $\rho_0$, the $\rho_{\rm gas}(R_{\rm vir})$ solved first increases then decreases.
    Right panel: the $\rho_{\rm gas}(R_{\rm vir})$ solved as a function of $\rho_0$ in diffrent halo masses.
    The solution of $\rho_0$ is determined from the left intersection of $\rho_{\rm ext}$ and $\rho_{\rm gas}(R_{\rm vir})$ curves.
    The local maximum of $\rho_{\rm gas}(R_{\rm vir})$ decreases with increasing halo mass,
    thus a critical $M_{\rm h, crit}$ exists above which no solution of $\rho_0$ can be found.
    In this case, $M_{\rm h,crit}$ lies between $6\times 10^6$ and $8\times 10^6~\msun$.
    }
    \label{fig:LE}
    \end{center}
\end{figure*}

In the Appendix, we briefly describe the method of how to calculate the critical gas density at the center
by solving the hydrostatic equation for an isothermal gas cloud (Eq.~\ref{eq:LE_iso}), where the gas 
pressure gradient force is balanced with the gas self-gravity and DM gravitational force.
For demonstration purpose, in the left panel of Fig.~\ref{fig:LE}, we show the radial profiles of gas with 
an effective sound speed of $c_{\rm eff} = 8.3 \mathrm{~km~s^{-1}}$ (corresponding to $T=10^4~\K$ gas in the absence of turbulence) 
for different values of $\rho_0$ in a DM halo with $M_{\mathrm{h}}=6\times10^6~\msun$ at $z=30$.
As the central density increases, the density at the virial radius $\rho_{\rm gas} (R_{\rm vir})$ does not increase monotonically
but has a local maximum value around $\rho_0 \simeq  10^{-21}~\gcc$.
In general, the maximum value of $\rho_{\rm gas} (R_{\rm vir})$ can be found for a given combination of $M_\mathrm{h}$, $z$, and $c_{\rm eff}$.
In the right panel of Fig.~\ref{fig:LE}, we present the relation between $\rho_{\rm gas} (R_{\rm vir})$ and $\rho_0$ for different halo masses 
($z=30$ and $c_{\rm eff} = 8.3 \mathrm{~km~s^{-1}}$ are fixed).
As seen in the left panel, each curve has a local maximum and the maximum value decreases with $M_{\rm h}$.
The density value at the outer boundary ($\rho_{\rm ext} = f_{\rm b} \rho_{\rm DM}$) weakly depends on $M_{\rm h}$ and $z$ 
through the concentration factor $c_{\rm vir}$, i.e., the three halos have $\rho_{\rm ext} \simeq 8\times 10^{-25}~\cc$, varying within $3\%$.
For $M_{\rm h}=6\times 10^6~\msun$, there exist two solutions where the boundary conditions are satisfied.
Since the solution with the higher value of $\rho_0$ is not stable, we adopt the solution with the lower value of $\rho_0$ 
\citep[see ][]{1955ZA.....37..217E,1958MNRAS.118..523B,LyndenBell_Wood_1968}. 
As the halo mass increases to $M_{\rm h}=8\times 10^6$ and $10^7~\msun$ , there is no hydrostatic solution of the gas cloud.
In our semi-analytical model, we calculate the hydro-static density profile which satisfies the boundary conditions at each time step
and quantify the critical halo mass $M_{\rm h, crit}$ above which the gas begins to collapse.
We note that this method can be applied to a wide range of $c_{\rm eff}$ and $z$ of interest in our paper.

\bibliography{ref}{}
\bibliographystyle{aasjournal}

\begin{table}
    \begin{center}
    \caption{Chemical Reactions}
    \begin{tabular}{l c r}
    \hline
    Number & Reaction & Reference \\ \hline \hline
     & H collisional reactions & \\
    \hline
    1  &   H        +     e$^-$ $\rightarrow $ H$^+$ +      2e$^-$   &  1\\
    2  &   H$^+$  +     e$^-$ $\rightarrow $ H       +    $\gamma $  &  2$^{\ast}$\\
    3  &   H        +     e$^-$ $\rightarrow $ H$^-$ +    $\gamma $&  3\\
    4  &   H$^-$  +    H       $\rightarrow $  H$_2$ +          e       &  4\\
    5  &   H        +    H$^+$ $\rightarrow $  H$_2^+$ + $\gamma $&  5\\
    6  &   H$_2^+$+   H        $\rightarrow $ H$_2$ +       H$^+$    &  6\\
    7  &   H$_2$  +    H        $\rightarrow $ 3H                          &  7\\
    8  &   H$_2$  +     H$^+$ $\rightarrow $ H$_2^+$ +       H       &  8\\
    9  &   H$_2$  +     e$^-$ $\rightarrow $ 2H      +     e$^-$      &  9\\
    10 &  H$^-$  +     e$^-$ $\rightarrow $ H        +     2e$^-$    & 10\\
    11 &  H$^-$  +     H$^+$ $\rightarrow $ 2H                          &  11\\
    12 &  H$^-$  +     H$^+$ $\rightarrow $ H$_2^+$ +   e$^-$     &  12\\
    13 &  H$_2^+$ +   e$^-$ $\rightarrow $ 2H                          &  13\\
    14 &  H$_2^+$ + H$^-$  $\rightarrow $  H$_2$ +      H           &  14\\
    15 &  3H                     $\rightarrow $ H$_2$   +     H           &  15\\
    16 &  2H       +   H$_2$ $\rightarrow $  2H$_2$                    &  16\\
    17 &  2H$_2$              $\rightarrow $  2H       +     H$_2$     &  17\\
    18 &  H$^-$   +   H  $\rightarrow $ 2H   +  e$^-$                  &  18\\
    19 &  H$^-$   +   H$_2^+$  $\rightarrow $ 3H                       &  19\\
    20 &  H$_2$   +   e$^-$  $\rightarrow $  H$^-$  +  H               &  20\\
    \hline
     & photo-dissociation and detatchment reactions & \\ 
    \hline
    21 &  H$_2$  + $\gamma $ $\rightarrow $  2H                     &  21\\
    22 &  H$^-$  + $\gamma $ $\rightarrow $ H    +      e$^-$    &  22\\
    23 &  H$_2^+$   + $\gamma $ $\rightarrow $  H   +   H$^+$   &  23\\
    \hline
    & He reactions & \\
    \hline 
    24 &  He  +  e$^-$      $\rightarrow $   He$^+$       +    2e$^-$         &  24\\
    25 &  He$^+$  +  e$^-$      $\rightarrow $   He       +    $\gamma $       &  25\\
    26 &  He$^+$  +  e$^-$      $\rightarrow $   He$^{++}$       +    2e$^-$   &  26\\
    27 &  He$^{++}$  +  e$^-$      $\rightarrow $   He$^+$       +    H$^+$  +$\gamma $     &  27\\
    28 &  H$_2$   +  He        $\rightarrow $   2H   +   He                              &  28\\
    29 &  H$_2$   +   He$^+$      $\rightarrow $   He   +  H  +   H$^+$    &  29\\
    30 &  H$_2$  +   He$^+$     $\rightarrow $   H$_2^+$       +    He   &  30\\
    31 &  He$^+$  +  H          $\rightarrow $   He        +   H$^+$    & 31\\
    32 &  He   +   H$^+$        $\rightarrow $   He$^+$  +   H          &32\\
    33 &  He$^+$  +   H$^-$     $\rightarrow $   He   +   H             &33\\
    34 &  He   +   H$^-$        $\rightarrow $   He   +   H   +   e$^-$  &34\\
    35 &  2H  +  He        $\rightarrow $   H$_2$  +  He            &35\\
    \hline

\end{tabular}
\label{table:reactions}
\end{center}
(1) \cite{1997NewA....2..181A}; (2) \cite{1992ApJ...387...95F}, Case B; (3) \cite{2017JPhB...50k4001M}; 
(4) \cite{2010Sci...329...69K}; (5) \cite{2011ApJS..193....7C}; (6) \cite{1979JChPh..70.2877K}; 
(7) \cite{1986ApJ...302..585M, 1983ApJ...270..578L}; (8) \cite{2004ApJ...606L.167S, 2011ApJS..193....7C}; 
(9) \cite{2002PPCF...44.1263T}; (10) \cite{Janev1987}; (11) \cite{1999MNRAS.304..327C}; 
(12) \cite{1978JPhB...11L.671P}; (13) \cite{1994ApJ...424..983S}; (14) \cite{1987IAUS..120..109D}; 
(15) \cite{2002Sci...295...93A, 1987JChPh..87..314O}; (16) \cite{1967JChPh..47...54J}; 
(17) \cite{1998ApJ...499..793M, 1987ApJ...318...32S} (18) \cite{Janev1987}; (19) \cite{1987IAUS..120..109D}; 
(20) \cite{1967PhRv..158...25S}; (21) \cite{Wolcott-Green_Haiman2011}; (22) \cite{2017JPhB...50k4001M}; 
(23) \cite{1994ApJ...430..360S}; (24) \cite{Janev1987}; (25) \cite{1998MNRAS.297.1073H}; (26) \cite{Janev1987}; 
(27) \cite{1992ApJ...387...95F}; (28) \cite{1987ApJ...318..379D}; (29) \cite{1984PhDT.......142B}; 
(30) \cite{1984PhDT.......142B}; (31) \cite{1989PhRvA..40.2340Z}; (33) \cite{1993ApJ...405..801K}; 
(33) \cite{1994JPhB...27.2551P}; (34) \cite{1982JPhB...15..951H}; (35) \cite{WALKAUSKAS1975691}; 
\end{table}

\end{CJK*}

\end{document}